\documentclass[%
 reprint,
 amsmath,amssymb,
 aps,
]{revtex4-2}

\usepackage{graphicx}
\usepackage{epstopdf}
\usepackage{subfigure}
\usepackage{hyperref}
\usepackage{dcolumn}
\usepackage{bm}
\usepackage{color}
\usepackage{url}
\newcommand\scalemath[2]{\scalebox{#1}{\mbox{\ensuremath{\displaystyle #2}}}}
\newcommand{\half}{\frac{1}{2}}
\newcommand{\Bo}[1]{\mathbf{{#1}}}
\newcommand{\MagAmp}{b}
\newcommand{\MagConic}{h}

\newcommand{\MagFieldConic}{\Bo{B}(t)=\MagAmp (\sin(\Omega t)\hat{\Bo{x}}+\cos(\Omega t)\hat{\Bo{y}}+\MagConic \hat{\Bo{z}})}

\begin{document}

\preprint{APS/123-QED}

\title{Spatial dynamics of flexible nano-swimmers under a rotating magnetic field}

\author{Chapnik Zvi}
 \altaffiliation{zvic@campus.technion.ac.il}
\author{Or Yizhar}%
 \email{izi@me.technion.ac.il}
\affiliation{%
Faculty of Mechanical Engineering, Technion – Israel Institute of Technology, Technion City, Haifa 3200003, Israel
}

\date{\today}

\begin{abstract}
Micro-nano-robotic swimmers have promising potential for future biomedical tasks such as targeted drug delivery and minimally-invasive diagnosis. An efficient method for controlled actuation of such nano-swimmers is applying a rotating external magnetic field, resulting in helical corkscrew-like locomotion. In previous joint work, we presented fabrication and actuation of a simple magnetic nano-swimmer composed of two nano-rods connected by a short elastic hinge. Experiments under different actuation frequencies result in different motion regimes. At low frequencies, in-plane tumbling; at higher frequencies, moving forward in a spatial helical path in synchrony with the rotating magnetic field; in further frequency increase, asynchronous swimming is obtained. In this work, we present mathematical analysis of this nano-swimmer motion. We consider a simple two-link model and explicitly formulate and analyze its nonlinear dynamic equations, and reduce them to a simpler time-invariant system. For the first time, we obtain explicit analytic solutions of synchronous motion under simplifying assumptions, for both solutions of in-plane tumbling and spatial helical swimming. We conduct stability analysis of the solutions, presenting stability transitions and bifurcations for the different solution branches. Furthermore, we present analysis of the influence of additional effects, as well as parametric optimization of the swimmer's speed. The results of our theoretical study are essential for understanding the nonlinear dynamics of experimental magnetic nano-swimmers for biomedical applications, and conducting practical optimization of their performance.

\end{abstract}

\maketitle

\section{\label{sec:level1}Introduction}

Nano-swimmers are of great interest due to their promising potential for biomedical purposes, with applications ranging from internal sensing to targeted drug delivery and even carrying out medical procedures \cite{li2017micro}. In order to understand how to create propulsion in the nanoscale dimensions, where motion is dominated by viscosity, one first needs to examine microorganisms’ motion. Microorganisms such as Salmonella, Helicobacter pylory, or Euglena gracilis are propelled by corkscrew motion of their tail \cite{crenshaw1996new}, and theoretical models have been developed \cite{purcell1977life} to investigate their motion in the dominant viscosity regime. The model of a sphere with a tail shows two types of locomotion mechanisms. First, rotation of a helical propeller rigid tail creating corkscrew forward motion; second, planar undulation of a flexible tail creating a forward motion.\\
In order to fabricate an artificial nano-machine capable of achieving those two types of locomotion, there is a need for actuation. Internal actuation in this small scale is not yet applicable mechanisms and would increase the manufacturing cost and complexity of the nanomachine. Thus, external actuation is an easier, practical solution. Using ultrasound, optic, electrical or magnetic forcing to create actuation torque were shown to be applicable \cite{balk2014kilohertz,chen2018small,xu2017fuel}. \\
The most common way is applying a time-varying external magnetic field while the nano-swimmer's structure contains a magnetic part, resulting in application of a magentic torque actuating the swimmer. Experiments showed that applying a rotating magnetic field on rigid nano-helix with a magnetic head creates forward locomotion \cite{dreyfus2005microscopic,zhang2009artificial}. However, fabricating such chiral/helical three-dimensional structure in nanoscale is a long and delicate procedure \cite{wang2015fabrication}. In recent years, new kind of flexible nano-swimmers has been fabricated \cite{gao2010magnetically,jang2018fabrication}, using rigid links connected by a short flexible hinge. Those nano-swimmers are much easier to manufacture. Actuated by a rotating magnetic field, it has been shown \cite{gao2010magnetically} that such nano-swimmers can also create spatial helical forward locomotion.  
Such magnetic nano-swimmers display different types of motion under different ranges of rotational frequencies of the magnetic field. At low-frequency, wobble in a plane without net swimming, and in higher frequencies swimming forward with spatial helical corkscrew motion \cite{gao2010magnetically}. At even higher frequencies, the nano-swimmer's motion loses its synchronization with the magnetic field's rotation, an effect named step-out \cite{SteOutishiyama2001swimming}.\\
In order to investigate these phenomena, there is a need to devise a simple mathematical model, which is amenable to explicit analysis. Some models have been developed for a single rigid body, starting from simple magnetic rigid ellipsoid under constant external torque \cite{ghosh2013analytical}, showing frequency-dependent stability transitions. Later, a model of a magnetic rigid chiral helix swimmer \cite{morozov2014chiral} shows frequency-dependent motion regimes, their stability, transitions, and step-out frequency for the loss of the synchronous motion.  Other works have considered simple magnetized micro-propellers made of aggregated beads, which were analyzed both experimentally \cite{cheang2014minimal,cheang2016versatile} and theoretically \cite{morozov2017dynamics}.\\
For the flexible nano-swimmer under a rotating magnetic field, experimental results \cite{WangEXgao2012cargo} followed by elastohydrodynamic model taking into account the elasticity of the nanowire and its hydrodynamic interaction with the fluid medium \cite{pak2011high}, considered a model with a rigid head and a flexible tail nano-swimmer, showing spatial motion with agreement to the theoretical model. Another work \cite{mirzae2020modeling} presented a multiple bead-spring model, allowing for arbitrary filament geometry and flexibility, and taking into account hydrodynamic interactions. This numerical work of multiple degrees-of-freedom (DOF) model also showed the motion dependence on frequency. In the case of a planar oscillating magnetic field, the work \cite{gutman2014simple} proposed a simple two-link model with a passive revolute joint acted by torsion spring under a planar oscillating magnetic field. Using this simple model allows finding analytically the motion's dependence on the oscillation frequency, including optimal swimming speed, as well as stability transitions \cite{harduf2018nonlinear} and bifurcations \cite{paul2023nonlinear}. 
For the case of a rotating magnetic field,  our collaborative work with Wu et al \cite{wu2021helical} 
considered the same two-link model in order to study the behavior of the experimental nano-swimmer. The experiments in \cite{wu2021helical} showed transitions between different motion regimes upon varying the rotational frequency of the magnetic field, which were captured by our  numerical simulations of the two-link model. However, an analytical investigation of the solution regimes, stability analysis, and performance optimization have not been presented in \cite{wu2021helical}.

The goal of this work is to complement the experimental results and numerical simulations in \cite{wu2021helical} by presenting full analytical treatment of the two-link model dynamics. We obtain explicit expressions for the frequency-dependent solutions and their transitions, and conduct numerical optimization of the nano-swimmer's performance with respect to various parameters,  namely, the head-tail links' length ratio, added axial component of the magnetic field, the nano-swimmer's bending stiffness, and the magnetic field's magnitude.

\section{Problem formulation}
We now present the two-link model and formulation of its dynamic equations of motion, as derived previously in \cite{wu2021helical}. 
\begin{figure}
	\centering
	\includegraphics[width=1\linewidth]{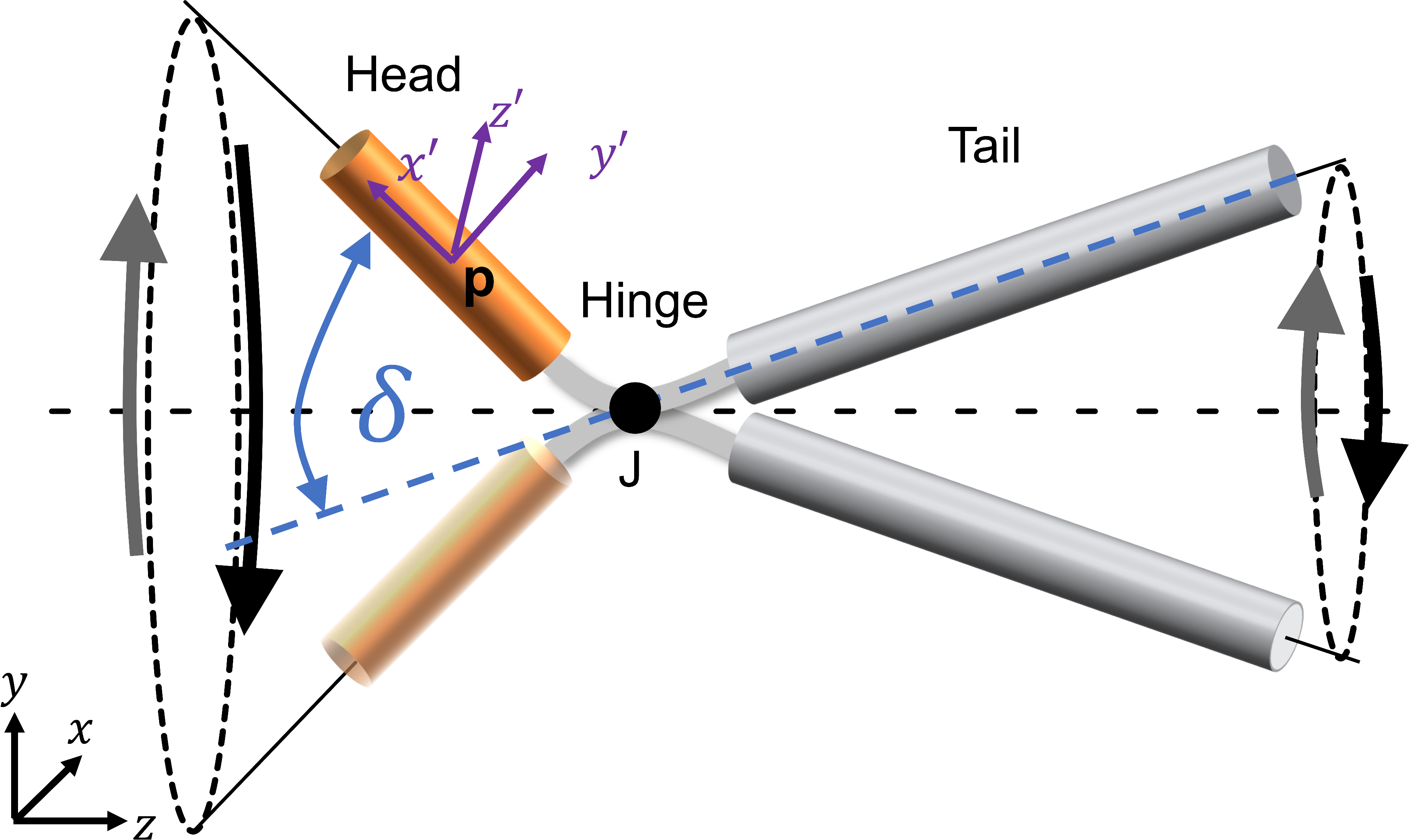}
	\caption{Schematic sketch of the two-link nano-swimmer model. p is the center of the head link, J is the pointed revolute joint point that connects the head and tail link. The joint's axis of rotation is $\Bo{z'}$, whereas the swimmers' links lie within $\Bo{x'}-\Bo{y'}$ plane.}
	\label{fig:model}
\end{figure}
The nano-swimmer's simplified model from \cite{wu2021helical}, as shown in Fig. \ref{fig:model}, consists of two rigid links: the magnetic link, named ``head" and the nonmagnetic link, named ``tail". A short thin nano-wire connects them, acting as a flexible hinge. The nano-swimmer moves in a 3D environment, where the position of the head link's center is denoted by coordinates $x,y,z$. In order to describe its orientation, we assign  a moving reference frame  $\{\hat{\Bo{x}}',\hat{\Bo{y}}',\hat{\Bo{z}}'\}$ fixed to the center of the head link, whose orientation with respect to the world frame is parametrized by three Euler angles $\phi,\theta,\psi$. The short flexible hinge is represented by a pointed uniaxial revolute joint along a common normal to the longitudinal axes of the head and tail links, which are assumed to lie on a common plane. The relative angle of this joint is denoted by $\delta$, as shown in Figure \ref{fig:model}. Note that this  highly idealized and simplified two-link model neglects the continuous bending deformation along the hinge wire and lumps it into a single elastic joint. Moreover, it neglects hydrodynamic interaction between the links, as we shall see below in equations \eqref{eq:f_hyd} and \eqref{eq:tau_hyd}. Nevertheless, it has been shown in \cite{wu2021helical} that this model is able to preserve the main features of the nano-swimmer's dynamical behavior.

The nano-swimmer's motion is described by a vector of 7 coordinates,  as:
\begin{equation} \label{eq.q} \Bo{q}=(x,y,z,\phi,\theta,\psi,\delta)^T.
\end{equation} 
The matrix describing the orientation of the rotating reference frame fixed to the head  $\{\hat{\Bo{x}}',\hat{\Bo{y}}',\hat{\Bo{z}}'\}$ with respect to the world reference frame $\{\hat{\Bo{x}},\hat{\Bo{y}},\hat{\Bo{z}}\}$, is obtained as the product of three extrinsic rotation matrices according to the Euler ZXZ convention as 
 $\Bo{R}_1=\Bo{R}_\phi\Bo{R}_\theta\Bo{R}_\psi$.\\
\begin{align}
	      \Bo{R}_1=  \left(\begin{array}{ccc} C_{\phi} \,C_{\psi}\text{-}C_{\theta}\,S_{\phi}\,S_{\psi} & \text{-}C_{\phi} \,S_{\psi}\text{-}C_{\psi}\,C_{\theta}\,S_{\phi} & S_{\phi}\,S_{\theta}\\ C_{\psi}\,S_{\phi}\text{+}C_{\phi} \,C_{\theta}\,S_{\psi} & C_{\phi} \,C_{\psi}\,C_{\theta}\text{-}S_{\phi}\,S_{\psi} & \text{-}C_{\phi} \,S_{\theta}\\ S_{\psi}\,S_{\theta} & C_{\psi}\,S_{\theta} & C_{\theta} \end{array}\right),
\label{eq:Rot1}
	\end{align}
 where $C_\phi \!=\! \cos\phi,C_\psi \!=\! \cos\psi,C_\theta \!=\! \cos\theta,S_\phi \!=\! \sin\phi, S_\psi\!=\! \sin\psi,S_\theta \!=\!\sin\theta$.
 In order to have a unique description for each orientation, we define the ranges of Euler angles as: $-\pi\le \phi<\pi$, $0<\theta\le \pi$, and $0\le \psi<2\pi$.\\
In order to formulate the dynamics of the system, we now describe the forces and torques acting on the nano-swimmer. First, we consider the hydrodynamic forces and torques. We assume that each link is axisymmetric about its longitudinal axis $\hat{\Bo{t}}_i$, $i=1,2$. Assuming Stokes flow and neglecting hydrodynamic interaction between the links, the drag force  $\Bo{\textbf{{f}}}^h_i$ and torque  $\boldsymbol{\tau}^h_i$ acting upon  link $i$ are linearly proportional to the velocity $\Bo{v}_i$ of the link's center and its angular velocity vector $\boldsymbol{\omega}_i$ \cite{happel2012lowSPCO1}. The hydrodynamic force and torque can thus be expressed as:
\begin{gather}
\label{eq:f_hyd} \Bo{\textbf{{f}}}^h_i=-c_t^i(\Bo{v}_i\cdot \hat{\Bo{t}}_i)\hat{\Bo{t}}_i-c_n^i (\Bo{v}_i-(\Bo{v}_i\cdot\hat{\Bo{t}}_{\color{black} i})\hat{\Bo{t}}_i)=\\ \left((c_n^i-c_t^i) \hat{\Bo{t}}_i\cdot\hat{\Bo{t}}_i^T-c_n^i\Bo{I}_{3{\color{black} \times}3} \right)\Bo{v}_i \nonumber \\
\label{eq:tau_hyd} \boldsymbol{\tau}^h_i=-d_t^i(\boldsymbol{\omega}_i\cdot \hat{\Bo{t}}_i)\hat{\Bo{t}}_i-d_n^i (\boldsymbol{\omega}_i-({\color{black}\boldsymbol{\omega}_i}\cdot\hat{\Bo{t}}_{\color{black} i})\hat{\Bo{t}}_i)=\\ \nonumber \left((d_n^i-d_t^i) \hat{\Bo{t}}_i\cdot\hat{\Bo{t}}_i^T-c_n^i\Bo{I}_{3{\color{black} \times}3} \right)\boldsymbol{\omega}_i
\end{gather}

where $c_{n}^i,c_{t}^i,d_n^i,d_t^i$ for $i=1,2$ 
are the linear and rotational drag coefficients along the normal and tangent to the longitudinal axis of the head and tail links. 
 In our simulations, we choose to model each link as a prolate spheroid with major radius $a_i=1/2 * L_i$ and minor radii $b_i$. The drag coefficients from \ref{eq:f_hyd},\ref{eq:tau_hyd} for such case were given in \cite{happel2012lowSPCO1,jeffery1922motionSFCO2}, and also detailed in \cite{Thesis}. Alternatively, one can use drag coefficients of finite slender cylindrical rods\cite{rodcoeef}. 
\\ 
Next, we describe the magnetic effects on the nano-swimmer. 
It is assumed that the magnetization vector of the head link is directed along its longitudinal axis $ \Bo{\hat{t}}_1$ and has magnitude $ m$. 
The external magnetic field is assumed to be spatially uniform and time varying, with rotating component in $\hat{x}-\hat{y}$ plane, and possibly an additional constant component in $\hat{\Bo{z}}$ direction. (Note the differences from \cite{wu2021helical} in the choice of the plane of rotation,  as well as the addition of constant component). That is, the magnetic field $\Bo{B}(t)$ rotates along the surface of a cone, and is formulated as:
\begin{gather}
\MagFieldConic
 \label{eq:MagField}
 \end{gather} 
The torque acting on the head link due to the rotating magnetic field is:
\begin{gather}\label{eq:MagTor}
 \boldsymbol{\tau}_m=m(\Bo{\hat{t}}_1\times \Bo{B}(t))
\end{gather} 

Lastly, we consider the nano-swimmer's elasticity.
The short flexible wire connecting the head and tail link is modeled as a  pointed uni-axial joint with torsion spring, applying torque:
\begin{gather}\label{eq:ElasticTo}
    \boldsymbol{\tau}_k=-k\delta \Bo{\hat z}'
\end{gather}

Due to the negligibility of inertial effects in Stokes flow, the swimmer's links are always in force and torque balance.  This includes the magnetic and elastic torques in \eqref{eq:MagTor}, \eqref{eq:ElasticTo}, as well as hydrodynamic forces and torques in \eqref{eq:f_hyd},\eqref{eq:tau_hyd}, which depend linearly on velocities, and thus can be expressed in terms of $\Bo{\dot{q}}$, where $\Bo{q}$ is the vector of generalized coordinates defined above in \eqref{eq.q}. Using kinematic relations,   (see detailed derivation in \cite{wu2021helical,Thesis}), one can write the balance equations and derive the nano-swimmer's dynamic equations of motion in the form of:
\begin{gather}
\Bo{A}(\Bo{q})\Bo{\dot{q}}=\Bo{b}(\Bo{q},t)
\label{eq:Aqb}
\end{gather}
 Detailed expressions of the square matrix $\Bo{A}(\Bo{q})$ and vector $\Bo{b}(\Bo{q},t)$ appear in \cite{Thesis}). Equation \eqref{eq:Aqb} gives a system of 7 coupled time-dependent nonlinear ordinary differential equations in $\Bo{q}(t)$, which can be integrated numerically. 

 It is important to note that the two links are modelled as axisymmetric bodies, so that their individual grand resistance matrices, when expressed in centered body-fixed frame aligned with principal symmetry axes, are diagonal matrices that contain no rotation-translation coupling terms. However, the connection of the two links with a rotational joint creates inter-link forces and moments enforcing the kinematic constraints on the links’ relative motion. When the two links are not aligned, i.e. $\delta \ne 0$, these internal forces and moments induce nonzero rotation-translation coupling in the overall resistance matrix. This coupling occurs even when neglecting hydrodynamic interaction between the links, which is a far-field effect that decays sharply with distance between the links’ boundary points. In order to illustrate this point, the Appendix shows detailed derivation of the grand resistance matrix of the two links rigidly attached at relative angle $\delta$. In fact, our problem is  even more complicated since the angle $\delta(t)$ evolves dynamically due to balance with the torsion spring and all other drag-induced and magnetic forces and moments.

\section{Numerical Simulations}
\begin{figure}
    \centering
    \includegraphics[width=0.8\linewidth]{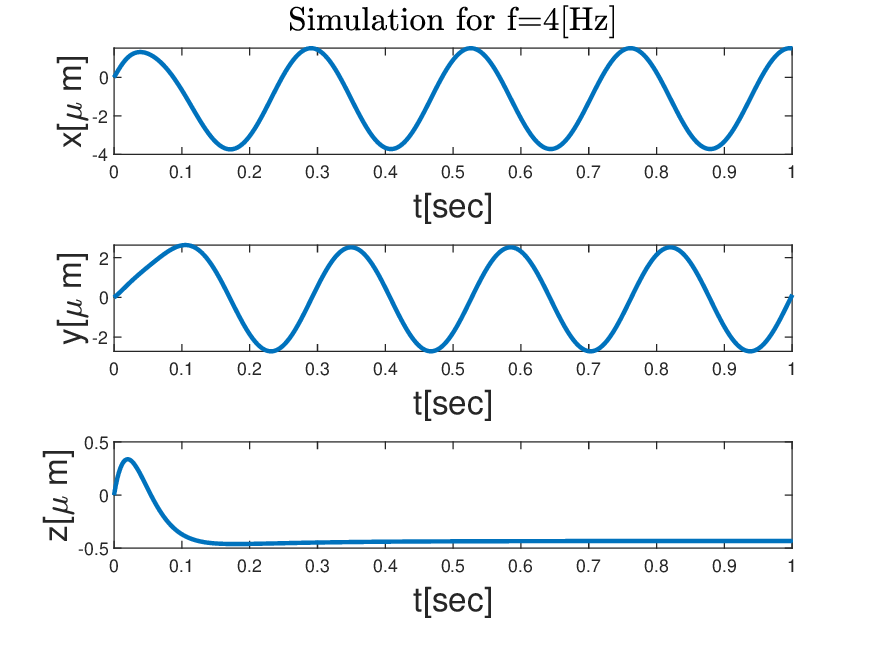}
        \includegraphics[width=0.8\linewidth]{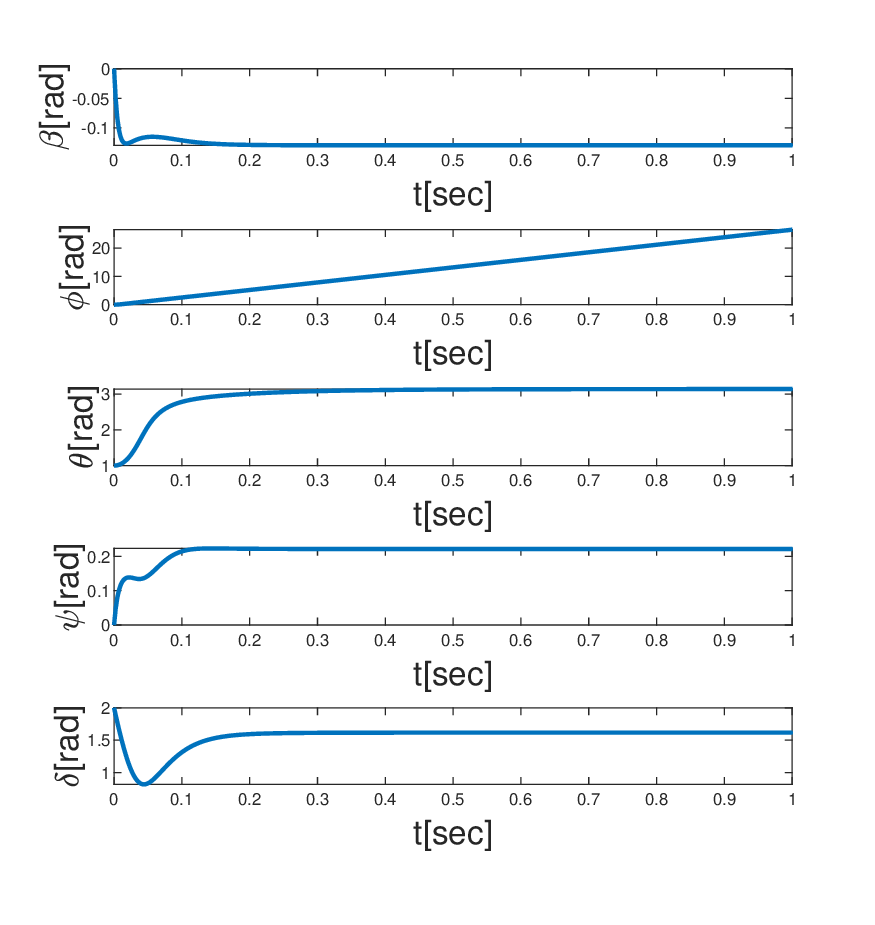}
    \caption{Numerical simulations under rotating magnetic field of f$=\Omega/2\pi=4[Hz]$, giving planar tumbling motion in $\hat{x}-\hat{y}$ plane.}
    \label{fig:tumb}
\end{figure}

\begin{figure}
    \centering
    \includegraphics[width=0.8\linewidth]{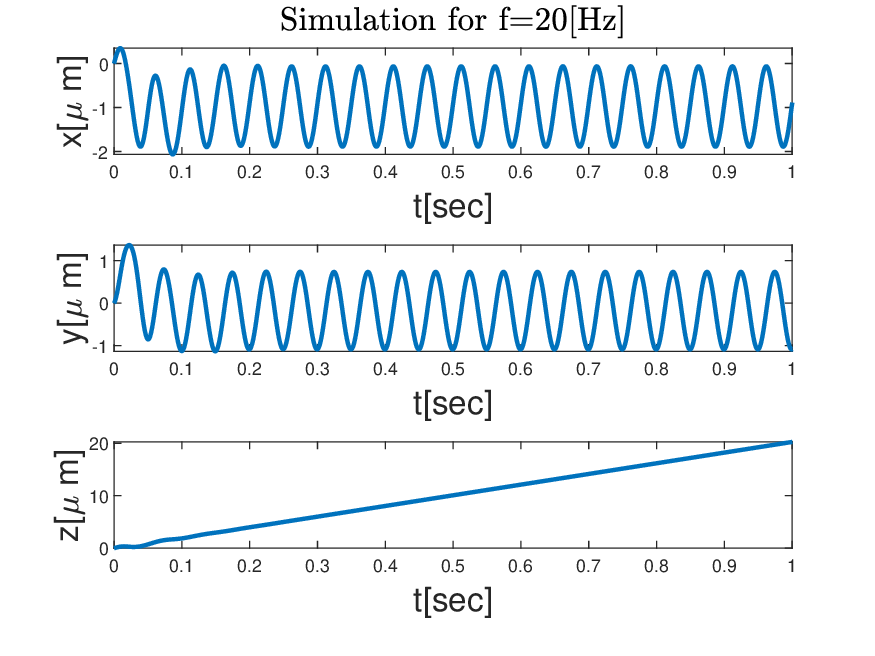} 
        \includegraphics[width=0.8\linewidth]{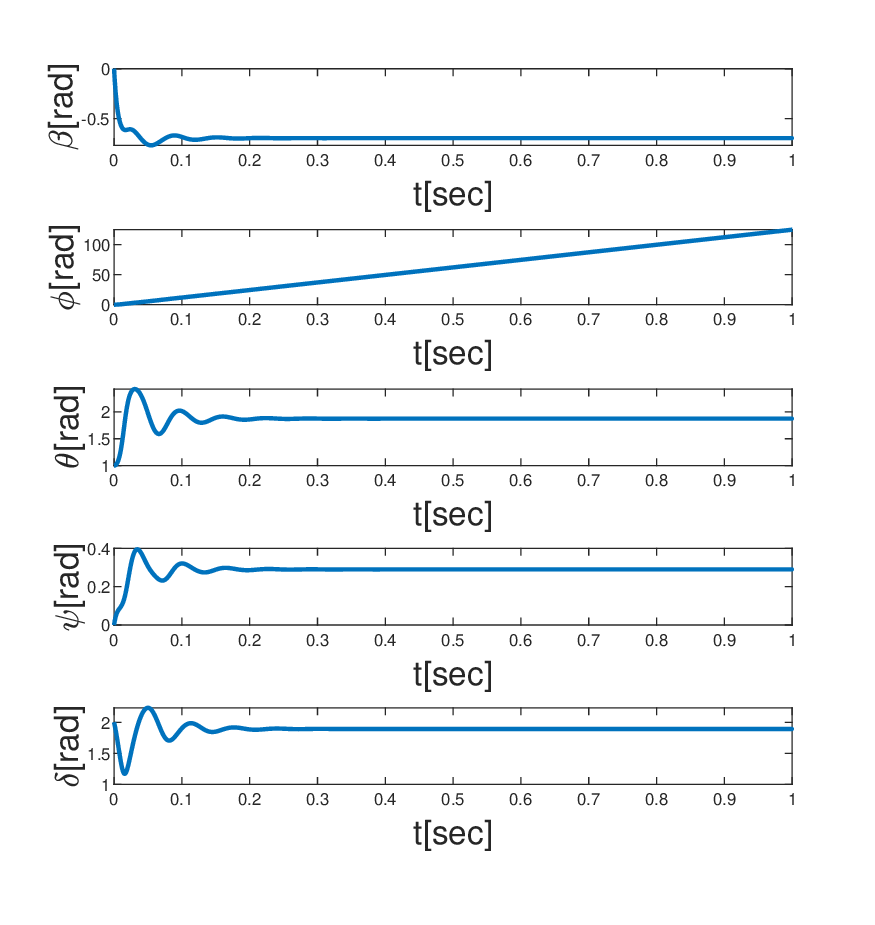 }
    \caption{Numerical simulations under rotating magnetic field of f$=\Omega/2\pi=20[Hz]$, giving synchronous motion of spatial helical trajectory.}
    \label{fig:Heli}
\end{figure}

\begin{figure}
    \includegraphics[ width=0.8\linewidth]{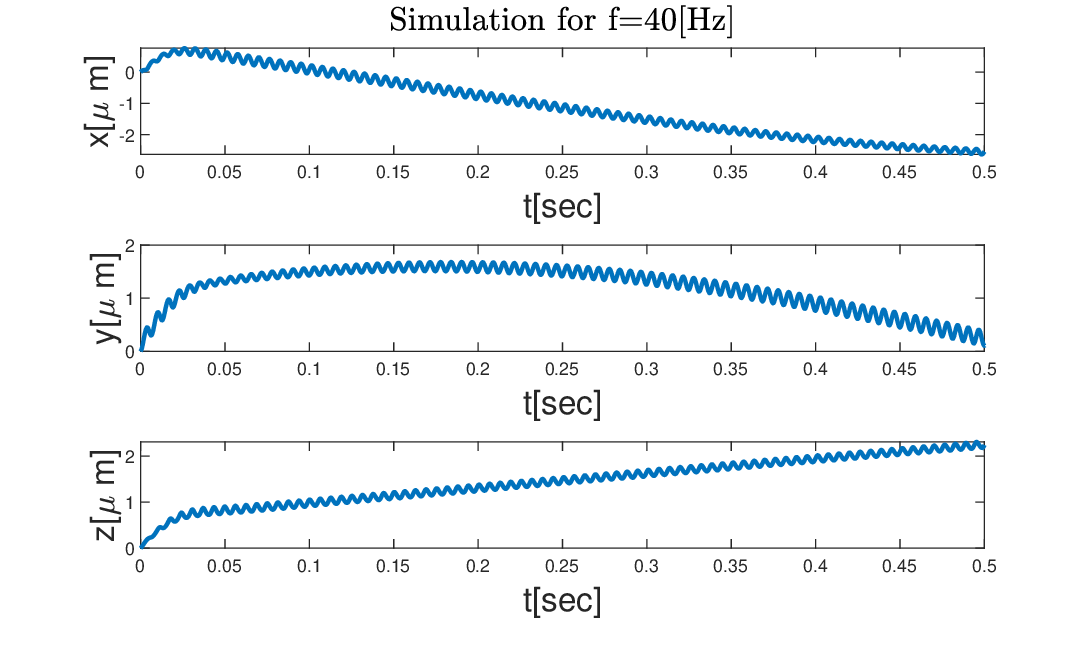}
        \includegraphics[width=0.8\linewidth]{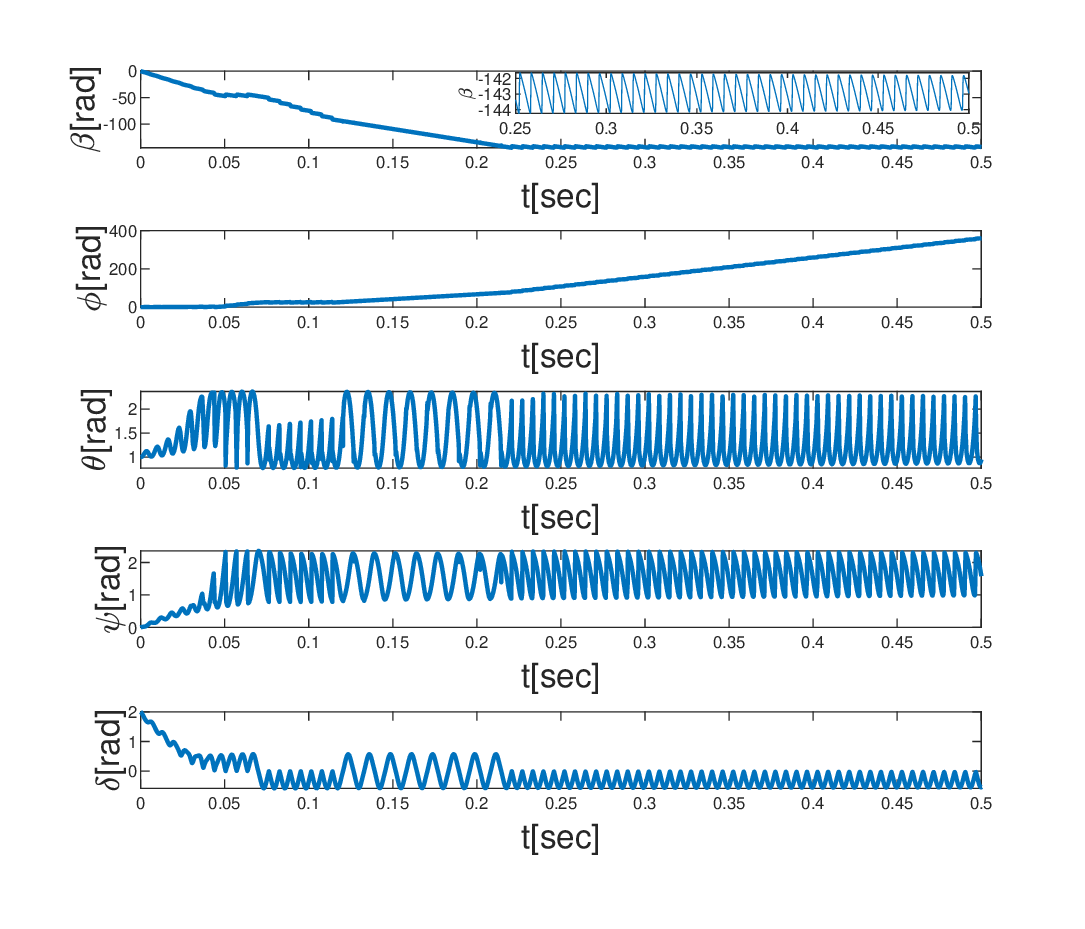}
    \caption{Numerical simulations under rotating magnetic field of f$=\Omega/2\pi=40[Hz]$, resulting in asynchronous motion.}
    \label{fig:Asin}
\end{figure}

In this section, we present numerical simulations representing the nano-swimmer's dynamics. We use the following physical values of parameters from \cite{wu2021helical}:
\begin{equation}
    \begin{split}
   &k=3.25\times 10^{-17} [N\cdot m],b=15 [mT],m=9000[A/m]\\
   &l_1=l_2=7.2 [\mu m],b_1=b_2=0.2[\mu m], \mu=0.0124 [N\cdot s/m^2] \label{eq:paraArt}
    \end{split}
\end{equation}
We integrate the equations of motion \eqref{eq:Aqb} using Matlab's $ode45$ function, assuming purely rotating  magnetic field with $\MagConic=0$, and using spheroid drag coefficients from \cite{happel2012lowSPCO1,jeffery1922motionSFCO2}. 
In all simulations, we consider initial conditions of orientation angles $\phi(0)=\psi(0)=0,\theta(0)=1 \;[rad]$ and joint angle $\delta(0)=2 \;[rad]$. By integrating in a wide range of actuation frequencies $\Omega[rad/sec]$, one  obtains three different types of motion: in low frequencies, planar tumbling - rigid rotation of the nano-swimmer in $\hat{x}-\hat{y}$ plane, in  synchrony with the rotating magnetic field, as shown in Figure \ref{fig:tumb}.
 After an initial transient, the angles $\theta(t),\psi(t),\delta(t)$  converge to constant steady-state values, whereas  $\phi(t)$ grows linearly with time $t$. In translational motion, $x(t)$ and $y(t)$ undergo periodic oscillations around zero, after a small initial transient, while $z(t)$ stays constant, resulting in zero net propulsion.
In higher frequencies, the nano-swimmer‘s body gets out of the $\hat{x}-\hat{y}$ plane and moves in a spatial corkscrew-like motion, following a helical path. After an initial transient, the steady-state motion is also synchronous with the applied magnetic field,  as shown in Figure \ref{fig:Heli}.
One can see that the three angles  $\theta(t),\psi(t),\delta(t)$ behave similarly to the previous case, and reach  constant values after an initial transient. In the translational motion, $x(t),y(t)$ are also oscillating about constant values, but $z(t)$ grows linearly in time, giving nonzero net propulsion along $\Bo{\hat{z}}$ direction in constant speed. \\
In higher frequencies as shown in Figure \ref{fig:Asin}, the nano-swimmer's motion loses its synchronization with the applied magnetic field, an effect known as “step-out” \cite{zhang2009characterizing}, the nano-swimmer continues moving forward in an asynchronous motion in all 7 coordinates.\\

\begin{figure}
	\centering
	\includegraphics[width=1\linewidth]{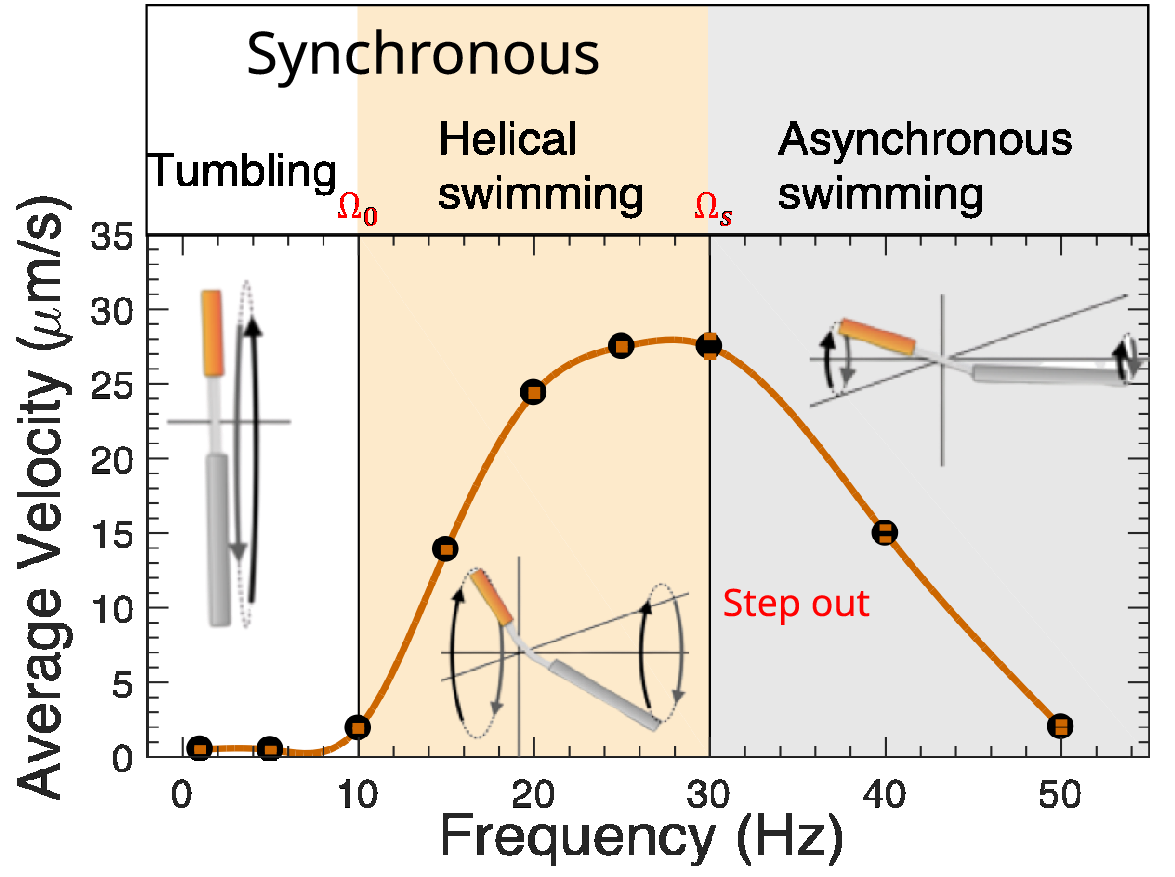}
	\caption{The average swimming speed $V_z$ as function of frequency. Solid line - calculated from numerical simulations; black dots are from experimental data \cite{wu2021helical}.}
	\label{fig:Speed}
\end{figure}

In Figure \ref{fig:Speed} we present the mean forward speed of the nano-swimmer in $\hat{\Bo{z}}$ direction, denoted by $V_z$, as a function of frequency $\Omega$ across the three different motion regimes.  
The mean speed changes smoothly as a function of the frequency. In the low frequency range of planar tumbling motion, the mean speed is zero.  
In higher frequencies, in the regime of spatial helical motion,  the nano-swimmer leaves the $\hat{x}-\hat{y}$ plane and starts a corkscrew locomotion. 
We define the critical frequency where the swimmer leaves the $\hat{x}-\hat{y}$ plane as $\Omega_0$. Increasing the frequency above $\Omega_0$, one can see an increase in the speed $V_z$ up to a maximum. Further increasing the frequency, the speed continues to decrease. Upon further increase of frequency, the swimmer's motions switch to the \textit{asynchronous} swimming, where the switching frequency is denoted by $\Omega_s$, also known as the \emph{step-out} frequency \cite{SteOutishiyama2001swimming,zhang2009characterizing}.  The circular markers in Figure \ref{fig:Speed} are values taken from experimental measurements in \cite{wu2021helical}.
\section{Analysis}
We now present analytical investigation of the two-link nano-swimmer dynamics. First, we reduce the equations of motion and make some simplifying assumptions that further reduce the number of parameters and simplify the equations. Then we study synchronous solutions, their bifurcations and stability transitions. 
\subsection{Reducing the Equations of Motion}
In order to investigate the system’s dynamic equations analytically, we now reduce the equations to a more compact form.
First, as shown in Figures \ref{fig:tumb},\ref{fig:Heli}, three of the coordinates $\theta,\psi,\delta$ are constant in steady state synchronous motion, but $\phi(t)$ is not. Since $\phi(t)$ maintains constant phase shift from the magnetic field orientation $\Omega t$, following \cite{morozov2014chiral},  we define a new coordinate of phase shift $\beta=\phi -\Omega t$ which in steady state is also constant. 
Using the new coordinates vector: $\Tilde{\Bo{q}}=(x,y,z,\beta,\theta,\psi,\delta)^T$, the system's equations of motion \eqref{eq:Aqb} are obtained as:\\
\begin{equation}
  \Tilde{\Bo{A}}(\Tilde{\Bo{q}})\dot{\Tilde{\Bo{q}}}=\Tilde{\Bo{b}}(\Tilde{\Bo{q}})  
  \label{eq:Aqtagb}
\end{equation}
Importantly, this new system is now time-invariant, and in the synchronous steady state, four coordinates converge to constant values, implying that it is an equilibrium state for those coordinates:
$\dot{\theta}=\dot{\beta}=\dot{\psi}=\dot{\delta}=0$. In Figures \ref{fig:tumb},\ref{fig:Heli}, the numerical simulations results for the coordinate $\phi$ and the coordinate $\beta=\phi -\Omega t$ are presented. One can see 
that $\beta(t)$ in both cases converges to constant values in steady state, after a transient response. \\
In order to further reduce the system's dynamic equations, we decompose the coordinates $\Tilde{\Bo{q}}$ into displacement part $\Bo{q}_d$ and angular part$,\Bo{q}_a$ as:
\begin{equation}
   \Tilde{\Bo{q}}=\left(\Bo{q}_d,\Bo{q}_a\right), \Bo{q}_d=\left(x,y,z\right)^T,\Bo{q}_a=\left(\beta,\theta,\psi,\delta\right)^T
\end{equation}
Then, we decompose the matrices  $\Bo{\Tilde{A},\Tilde{b}}$ from \eqref{eq:Aqtagb} into blocks as:

\begin{gather}
\underbrace{\left(\begin{matrix}
	\Bo{A}_{dd}&\Bo{A}_{da}\\
	\Bo{A}_{ad}& \Bo{A}_{aa}
	\end{matrix}\right)}_{\Bo{\Tilde{A}}(\Bo{q}_a)}
\left(\begin{matrix}
\Bo{\dot{q}}_d\\
\Bo{\dot{q}}_a
\end{matrix}\right)=\underbrace{\left(\begin{matrix}
\Bo{0}\\\Bo{g}
\end{matrix}\right)}_{\Bo{\Tilde{b}}(\Bo{q}_a)}
\label{eq:separateEQ}
\end{gather}

Note that equations \eqref{eq:separateEQ} depend only on the angles $\Bo{q}_a$ due to translational invariance. The translational velocity vector $\Bo{\dot{q}}_d$ can thus be extracted from \eqref{eq:separateEQ} as:
\begin{gather}
\dot{\Bo{q}}_d=-(\Bo{A}_{dd})^{-1}\Bo{A}_{da}\dot{\Bo{q}}_a
\label{eq:Velocity}
\end{gather}
By substituting \eqref{eq:Velocity} to \eqref{eq:separateEQ}, one obtains:
\begin{gather}
\underbrace{\left(
	\Bo{A}_{ad}\left(-(\Bo{A}_{dd})^{-1}\Bo{A}_{da}\right)+ \Bo{A}_{aa}\right)}_{\Bo{A'}(\Bo{q}_a)}\dot{\Bo{q}}_a
=\Bo{g}(\Bo{q}_a)
\label{eq:4DOF}
\end{gather}
Thus, the system is reduced to a 4-DOF time-invariant system in  $\Bo{q}_a$ only. Assuming $\Bo{A'}$ is an invertible matrix, one can solve the equilibrium equation $\Bo{\dot q}_a=0$  from the right side of equation \eqref{eq:4DOF} as:
\begin{equation}
     \Bo{g}(\Bo{q}_a)=\Bo{0}
     \label{eq:4DOFEQ}
\end{equation}
Solutions $\Bo{q}_a=\Bo{q}_e$ of equation \eqref{eq:4DOFEQ} correspond to steady-state synchronous motion.

\subsection{Simplifying Assumptions and Scaling} \label{Sec:SA}
The dynamic equations \eqref{eq:4DOF} depend on 15 physical constants, which calls for simplifications and reduction.
We now apply the following simplifying assumptions in order to reduce the number of parameters. First, we assume a planar magnetic field with no $\hat{\Bo{z}}$ component, $\MagConic=0$ in \eqref{eq:MagField}. Second, we consider equal links' lengths  $l_1=l_2=l$. (These two assumptions are relaxed later in Section \ref{sec.general_case}). Third,  drag coefficients are simplified, assuming slender prolate spheroids according to \cite{happel2012lowSPCO1,jeffery1922motionSFCO2}. We use the following assumptions to reduce the number of drag coefficients. The translational coefficients are approximated as  $c_t\approx \half c_n$. For the rotational drag coefficients, since slender bodies satisfy $d_t \ll d_n$, we neglect it and assume $d_t=0$. 
We also take $d_n= c_n l^2 /12$. These simplifications are fairly accurate approximation for slender spheroids \cite{happel2012lowSPCO1,jeffery1922motionSFCO2}.

{\bf Non-dimensional equations:}
First, we define $l$ as the characteristic length. Next, we define two characteristic time scales, $t_m$ the magnetic characteristic time and $t_k$ the stiffness characteristic time:
\begin{equation}
    t_m=\frac{c_{t}l^2}{\MagAmp m},t_k=\frac{c_{t}l^2}{k}\label{eq:time},
\end{equation}
Following \cite{gutman2014simple}, the ratio between these two time scales is defined as the nondimensional parameter $\alpha=t_m/t_k=k/\MagAmp m$. We scale the time in  \eqref{eq:4DOF} by the magnetic characteristic time  $t_m$, and define the nondimensional frequency $\omega=\Omega t_m  $. The scaled and reduced dynamic equations now have the form:
\begin{equation}
\Bo{A'}(\Bo{q}_a) \Bo{\dot{q}}_a
=\Bo{g}(\Bo{q}_a)
    \label{eq:SimpEQAqg}
\end{equation}
Where ``dot" now represents derivative with respect to the scaled time $t/t_m$. The matrix $\Bo{A'}(\Bo{q}_a)$ in \eqref{eq:SimpEQAqg} is given in Table \ref{tab.Atag}.
\begin{table*}[ht]
\centering
\[
\Bo{A'}(\Bo{q}_a) = 
\left(
\begin{array}{cccc}
\frac{1}{24}\left( 5 C_\psi +3 S_\delta S_\psi -5  c^2_\delta  C_\psi +5  C_\delta S_\delta S_\psi\right) & 
-\frac{S_\theta }{24}\left(3 C_\psi S_\delta -5 S_\psi +5 C_\delta^2 S_\psi +5 C_\delta C_\psi S_\delta\right) & 
0 & 0 \\
\frac{1}{12}\left( 3 S_{\delta - \psi} - 5 S_{2 \delta + \psi} - 9 S_{\delta + \psi} - 15 S_\psi \right) & 
\frac{1}{48} S_{\theta} \left(9 C_{\delta + \psi} + 15 C_{\psi} + 3 C_{\delta - \psi} + 5 C_{2 \delta + \psi} \right) & 
0 & 0 \\
0 & \frac{C_\theta (3+ 3 C_\delta)}{3(3 - C_\delta)} & \frac{3+ C_\delta}{3(3 - C_\delta)} & \frac{3+ C_\delta}{6(3 - C_\delta)} \\
0 & \frac{C_\theta (3+ C_\delta )}{6(3 - C_\delta)} & \frac{3+ C_\delta}{6(3 - C_\delta )} & \frac{9+c^2_\delta}{6(9 - c^2_\delta)}
\end{array}
\right)
\]
\caption{\label{tab.Atag} The matrix $\Bo{A'}(\Bo{q}_a)$.  The notations $S_\gamma$ and $C_\gamma$ are abbreviations of $\sin (\gamma)$ and $\cos (\gamma)$, respectively, for all angles $\gamma$}
\end{table*}
The vector $\Bo{g}(\Bo{q}_a)$ is given as: 
\begin{equation} \Bo{g}(\Bo{q}_a) = \left(
\begin{split}    &\omega \sin\theta \sin\delta \left(a_1(\delta )\sin(\psi )+b_1(\delta )\cos(\psi )\right)\\
    &\sin\theta \left(\sin\beta +\omega b_2(\delta )\cos(\psi )+a_2(\delta )\sin(\psi )\right)\\
    &\cos\beta \sin\psi +\cos\theta \left(\sin\beta \cos\psi +\omega a_3(\delta )\right)\\
    &\alpha\delta -\omega \cos\theta a_4(\delta )    \end{split} \right)
    \label{eq:SimEqua}
\end{equation}
where 
\begin{equation}
\begin{split} &a_1(\delta )=-\frac{5}{3}\sin(\delta )\\
&b_1(\delta )=\frac{5}{3}\cos(\delta )+1\\
&a_2(\delta )=-\frac{1}{4}\sin(\delta )\left(1+\frac{5}{3}\cos(\delta )\right)\\
&b_2(\delta )=\frac{1}{4}\left(\frac{5}{3}+\cos(\delta )\left(2+\frac{5}{3}\cos(\delta )\right)\right)\\
&a_3(\delta )=\frac{6+\cos(\delta )\left(4+\frac{1}{3}\cos(\delta )\right)}{9-\cos^2(\delta )}\\
&a_4(\delta )=-\half a_3(\delta )\end{split}
    \label{eq:SimEquaAiBi}
\end{equation}
Equation \eqref{eq:SimEqua} reached a compact simplified form, which enables solving the equilibrium equations \eqref{eq:4DOFEQ}. Note that this solution hold only if the matrix $\Bo{A'}(\Bo{q}_a)$ in \eqref{eq:SimpEQAqg} is invertible. From Table \ref{tab.Atag}, the determinant of $\Bo{A'}(\Bo{q}_a)$ can be obtained as $\mathrm{det}(\Bo{A'}(\Bo{q}_a))=\frac{1}{1296}\sin^2 \delta \sin \theta$. One case of singularity of 
$\Bo{A'}(\Bo{q}_a)$ 
is $\sin(\delta)=0$. This configuration describes the cases where the two-links are aligned $\Bo{\hat{t}}_1||\Bo{\hat{t}}_2$. 
The physical reason for this singularity is our simplifying assumption that $d_t^i=0$, which means that there is no drag resistance for rotation about the links’ longitudinal direction. When $d_t\ne0$, this singularity of $\Bo{A'}$ at $\sin(\delta)=0$ is removed. Nonetheless, we choose here to keep the simplifying assumption of $d_t=0$, since the case of $\sin(\delta)=0$ is anyway not an equilibrium solution of \eqref{eq:SimEqua} for $\omega \ne 0$.
The second case of singularity of $\Bo{A'}(\Bo{q}_a)$ is $\sin (\theta)=0$. This is associated with the swimmer lying in the $\hat{x}-\hat{y}$ plane - the planar tumbling motion, as discussed below.

\subsection{Planar Tumbling Solution}\label{sec:tumsol}
\begin{figure}
    \centering
    	\includegraphics[width=0.7\linewidth]{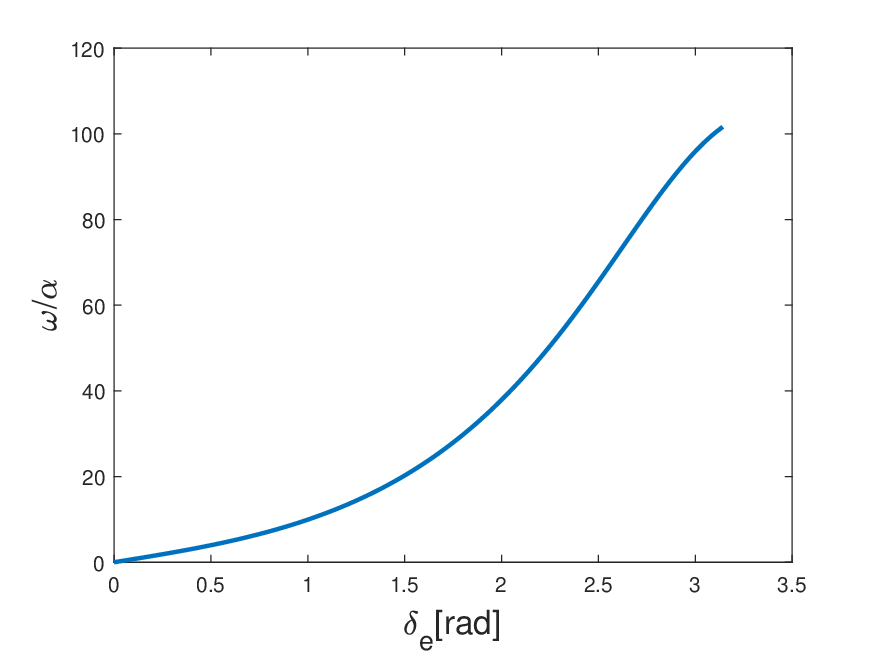}
    \caption{Nondimensional scaled frequency $\omega/\alpha$  as a function of equilibrium value of the link angle $\delta=\delta_e$ for the planar tumbling solution as obtained from \eqref{eq:TmOmegaDel}.}
    \label{fig:WtoDelPlanar}
\end{figure}

We now solve the equilibrium equations \eqref{eq:4DOFEQ} for planar tumbling motion.  In this motion regime, the nano-swimmer's links are fully contained in $\hat{x}-\hat{y}$ plane (i.e. $\theta=\{0,\pi\}$), and rotates as a rigid body with constant joint angle $\delta$, without making any progression in  $\Bo{\hat{z}}$ direction. This solution occurs at low frequencies $\omega$, and is represented by the simulation results in Figure \ref{fig:tumb}, and the left region in Figure \ref{fig:Speed}. The equilibrium values of $\Bo{q}_a$ will be denoted by $\Bo{q}_e=(\beta_e,\theta_e,\psi_e,\delta_e)$.  Looking at the first two terms in \eqref{eq:SimEqua},  one can easily observe that they both vanish for $\sin (\theta_e)=0$, meaning a planar solution.\\
\begin{equation}
    \sin \theta_e=0\rightarrow \theta_e=\pi \label{eq:TumSTheta}
\end{equation}
By substituting this solution into the other terms in \eqref{eq:SimEqua} and equating to zero, one obtains:
\begin{gather}
    \cos\beta_e\sin\psi_e -\left(\sin\beta_e\cos\psi_e+\omega a_3(\delta_e)\right)=0\label{eq:Tum1}\\
    \alpha\delta_e=-\omega a_4(\delta_e)=\omega\frac{6+\cos(\delta_e)\left(4+\frac{1}{3}\cos(\delta_e)\right)}{2\left(9-\cos^2(\delta_e\right))} \label{eq:Tum2} 
\end{gather}
Where $a_3(\delta_e),a_4(\delta_e)$ are given in \eqref{eq:SimEquaAiBi}. 
Equation \eqref{eq:Tum2} is a scalar function in $\delta_e, \omega$. For a given frequency $\omega$ one needs to solve a transcendental equation numerically for $\delta_e$. On the other hand, assuming given $\delta_e$ gives a simple analytical solution for $\omega$:\begin{equation}
    \omega=-\frac{\alpha \delta_e}{a_4(\delta_e)}= \frac{2\alpha \delta_e\left(9-\cos^2(\delta_e)\right)}{6+\cos(\delta_e)\left(4+\frac{1}{3}\cos(\delta_e)\right)} \label{eq:TmOmegaDel}
\end{equation}
Figure \ref{fig:WtoDelPlanar}   shows a plot of  the scaled nondimensional frequency $\omega/\alpha$ as a function of  $\delta_e$ within the range $[0,\pi]$.

We now proceed to solving the equilibrium values of the other angles $\psi,\beta$. Eq. \eqref{eq:Tum1}  can be rewritten as:
\begin{equation}
    \sin(\beta_e-\psi_e)=\sin(\gamma_e)=\omega a_3(\delta_e)\label{eq:TmbBpmP}
\end{equation}
This gives two solutions for $\gamma_e=\beta_e- \psi_e$ as a function of $\omega,\delta_e$.
Note that for $\sin (\theta_e)=0$, the first two terms in  \eqref{eq:SimEqua} vanish. Thus, one more equation is needed in order to find the angles $\beta_e$ and $\psi_e$ separately. This is resolved by considering the  ``full equilibrium equation" $\Bo{\dot{q}}_a=\Bo{A'}^{-1}\Bo{g}=\Bo{0}$. Interestingly, the singularity of the  matrix $\Bo{A'}$ for $\sin (\theta)=0$ is canceled in the multiplication $\Bo{A'}^{-1}\Bo{g}$, which becomes a well-defined finite vector. 
Substituting $\psi_e=\beta_e-\gamma_e$ into the second  element of  $\Bo{A'}^{-1}\Bo{g}=0$  gives an equation where the only unknown is $\beta_e$:

\begin{equation}
      f_1 \sin(\beta_e)^2 + f_2 \sin(\beta_e) \cos(\beta_e) +f_3=0,
      \label{eq:f123}
\end{equation}
   where:
\begin{gather}   
f_1=15 \alpha \cos(\gamma_e) \sin(\delta_e) \cos(\delta_e) - 15 \alpha \sin(\gamma_e) \cos(\delta_e)^2\nonumber\\   + 15 \alpha \sin(\gamma_e) + 9 \alpha \cos(\gamma_e) \sin(\delta_e)\\
f_2=15 \alpha \cos(\gamma_e) - 15 \alpha \sin(\delta_e) \sin(\gamma_e) \cos(\delta_e)\\  \nonumber- 15 \alpha \cos(\gamma_e) \cos(\delta_e)^2 - 9 \alpha \sin(\delta_e) \sin(\gamma_e)\\
f_3= 4 \omega - 4 \omega \cos(\delta_e)^2
\end{gather}
Since $\gamma_e$ is already solved in \eqref{eq:TmbBpmP} as a function of $\delta_e,\omega$, equation \eqref{eq:f123} is solved as:
\begin{align}
\begin{split}
\beta_e=\arctan\left(\frac{(\pm f_2^2 \mp \sqrt{f_2^4-4f_2^2f_3(f_1+f_3})f_4}{\pm f_2 f_3  f_4}\right)
\end{split}\label{eq:TmSolBeta}
\end{align}
Where
\begin{align}
f_4=\sqrt{\frac{f_2^2-2f_1f_3\pm \sqrt{f_2^4-4f_2^2f_3(f_1+f_3)}}{f_1^2+f_2^2}}
\end{align}
We obtain four different solution pairs $(\beta_e,\psi_e)$ for every choice of $\delta_e,\theta_e$. 
Each of these solutions has additional conditions for its existence, defined by the following inequalities. From \eqref{eq:TmbBpmP}, one obtains:
\begin{gather}
    |\omega a_3(\delta_e)|\le1
\end{gather}
From \eqref{eq:TmSolBeta}, one obtains:
\begin{gather}
   f_2^4-4f_2^2f_3(f_1+f_3)\ge 0\\
   f_4(\omega,\delta_e)\ge 0
\end{gather}
Those inequalities impose conditions for existence of each solution branch of planar tumbling.

\begin{figure}
	\includegraphics[width=0.7\linewidth]{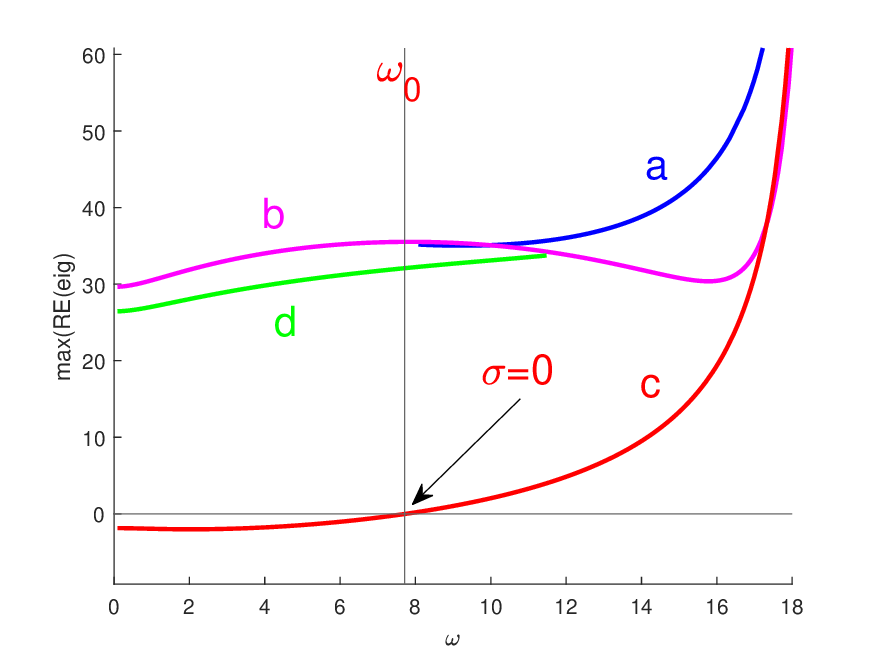}
	\caption{The quantity $max\{Re(\lambda_i({\Bo J}))\}$ as a function of nondimensional frequency  $\omega$, for the four solution branches of planar tumbling.}
		\label{fig:BetaEEigl}
\end{figure}

{\bf Stability analysis:}
We now examine the stability of steady-state solutions of planar tumbling using linear analysis, by the Jacobian matrix, calculated analytically.
\begin{gather}
    \Bo{J} = \left.\frac{d\left[ \Bo{A'}^{-1}\Bo{g} \right]}{d \Bo{q}}\right|_{\Bo{q}=\Bo{q}_e}
    = \begin{pmatrix}
     \sigma    & \Bo{0} \\
     \Bo{0}     & \Bo{J_m}
    \end{pmatrix}
    \label{eq:TmJac}
\end{gather}
For the planar tumbling solution, the resulting Jacobian matrix $\Bo{J}$ has a block-diagonal structure as in \eqref{eq:TmJac}. One eigenvalue of $\Bo{J}$ is the diagonal term $\sigma$, given by:
\begin{align}
    \sigma=-\frac{1}{(4(\cos(\delta_e)^2 - 1)}(3\alpha\sin(\beta_e)(3\cos(\psi_e)\sin(\delta_e)\nonumber \\ - 5\sin(\psi_e) + 5\cos(\delta_e)^2\sin(\psi_e) + 5\cos(\delta_e)\cos(\psi_e)\sin(\delta_e)))\label{eq.sigma}
\end{align}
The three other eigenvalues  $\Bo{J}$ belong to the $3\times3$ matrix block $\Bo{J_m}$. The Jacobian's expressions can be found at \cite{Thesis}.\\
\begin{figure}[b]
	\includegraphics[width=0.7\linewidth]{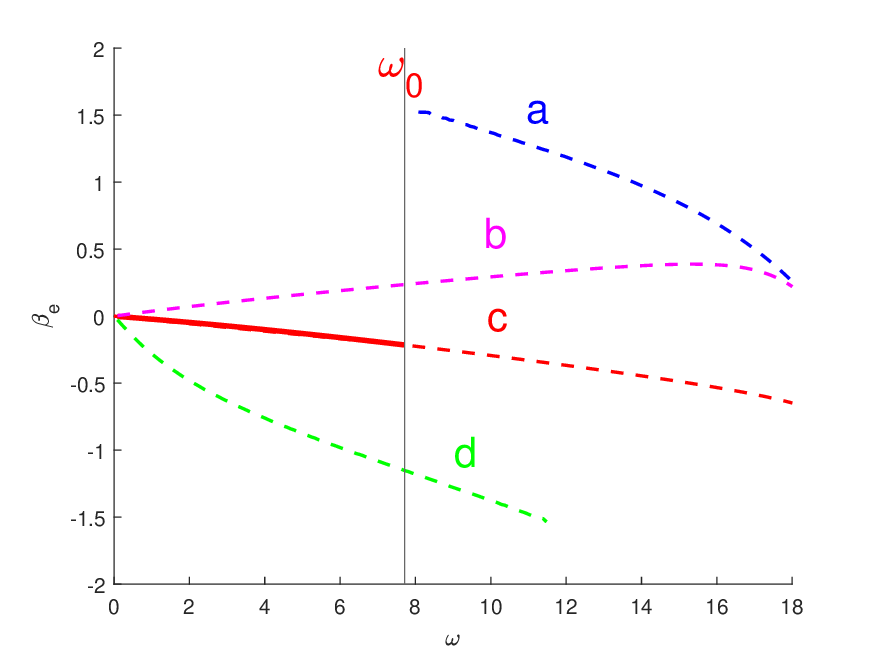}
	\caption{Multiple solutions branches of  $\beta_e$ for  planar tumbling, as a function of nondimensional frequency  $\omega$. Solid/dashed curves denote stable/unstable solutions, respectively.}
	\label{fig:BetaEsol}
\end{figure}
In order for the solution $\Bo{q}_e$ to be a stable equilibrium, all eigenvalues of $\Bo{J}$ in \eqref{eq:TmJac} are required to have negative real parts, $Re(\lambda_i(\Bo{J}))<0$. Thus, a necessary condition for stability is $\sigma<0$.  Figure \ref{fig:BetaEsol} shows four solutions branches of $\beta_e$ angle, denoted by $\{a,b,c,d\}$, as a function of the nondimensional frequency $\omega$. Solid lines denote stable solutions, while dashed lines denote unstable ones. There is only one branch that has a stable part, solution \{c\} in the figure. This can be verified by Figure \ref{fig:BetaEEigl}, which plots the maximal real part of eigenvalues of $\Bo{J}$ as a function of nondimensional frequency $\omega$.  For solution  
\{c\}, the eigenvalue that has the maximal real part is $\sigma$, and thus the sign of $\sigma$ determines its stability. One can find the critical value of  $\delta_e$ corresponding to stability transition from solving $\sigma=0$ in \eqref{eq.sigma} using the solution of $\psi_e(\delta_e),\beta_e(\delta_e)$, and then use \eqref{eq:TmOmegaDel} to find the critical frequency $\omega_0$. This frequency  where the planar tumbling solution loses its stability, also corresponds to a bifurcation point where the spatial helical solution emerges, as we shall see below.   That is, it is the frequency of transition between the two motion regimes, as illustrated in Figure \ref{fig:Speed}. Using the physical parameter values in \eqref{eq:paraArt}, the nondimensional transition frequency $\omega_0$ corresponds to physical frequency value of $\Omega=6.5 [Hz]$, which is reasonably comparable to the transition frequency estimated from  the nano-swimmer's experimental measurements in \cite{wu2021helical}, as  illustrated in Figure \ref{fig:Speed}.
In summary, we found a full analytical solution for the planar tumbling motion, as well as existence and stability conditions.
For known $\delta_e$ we obtained an explicit solution for $\omega$. On the other hand, For known $\omega$ we obtain a semi-analytic solution, solving $\delta_e$ numerically from a transcendental equation, and then solving the other coordinates analytically.

\begin{figure*}
	\includegraphics[width=0.8\linewidth]{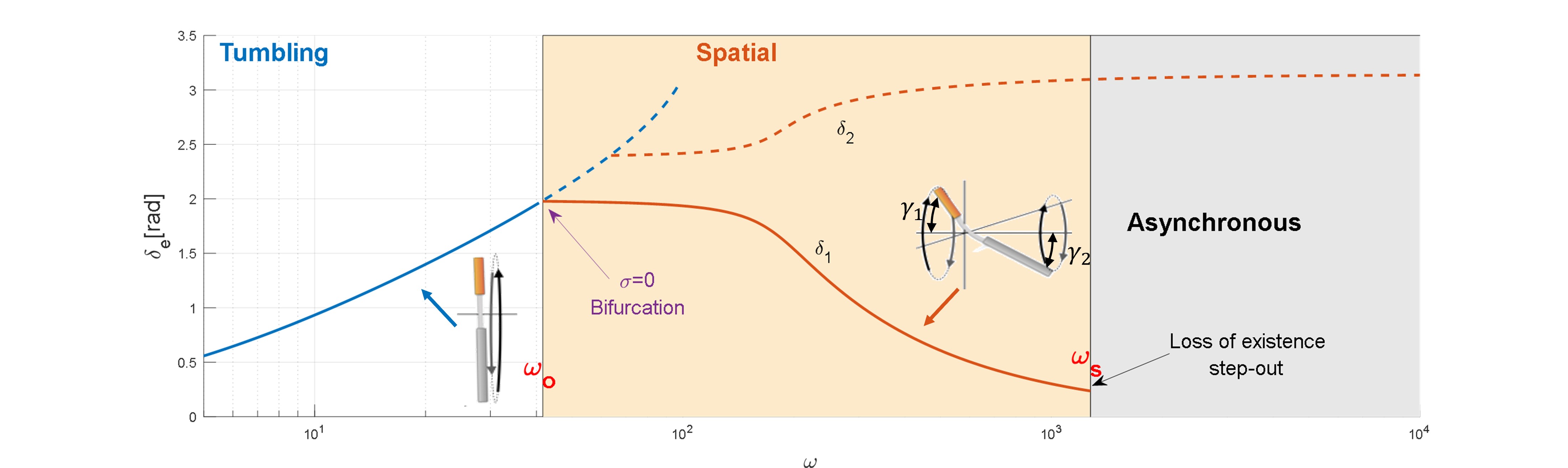}
	\caption{Synchronous solutions branches of the link angle $\delta_e$ as a function of nondimensional frequency  $\omega$. Blue curve - planar tumbling; red curves - two branches of spatial helical solution. Solid curves - stable solutions; dashed curves - unstable solutions.}
	\label{fig:DeltaStab}
\end{figure*}

\subsection{Spatial helical solution}\label{sec:helsol}
 We now consider the spatial helical solution, in which the nano-swimmer is oriented out of the $\hat{x}-\hat{y}$ plane (i.e. $\sin \theta_e \ne 0$), with constant joint angle $\delta$, and its translation is along a helical trajectory with net progression in $\Bo{\hat{z}}$ direction. This solution occurs at medium frequencies $\omega$, and is represented by the simulation results in Figure \ref{fig:Heli}, and the left region in Figure \ref{fig:Speed}. For obtaining this
solution, we revisit the equilibrium equations \eqref{eq:4DOFEQ} and \eqref{eq:SimEqua}. Assuming that $\sin\theta_e\ne0$, they can be simplified to:
\begin{gather}
     \omega \sin\delta \left(a_1(\delta)\sin(\psi)+b_1(\delta)\cos(\psi)\right)=0\label{eq:EqA}\\
   \sin\beta +\omega b_2(\delta)\cos(\psi)+a_2(\delta)\sin(\psi)=0\label{eq:EqB}\\
    \cos\beta\sin\psi +\cos\theta\left(\sin\beta\cos\psi+\omega a_3(\delta)\right)=0\label{eq:EqC}\\
    \alpha\delta=\omega \cos\theta a_4(\delta)  \label{eq:EqD} 
\end{gather}
where $a_i(\delta),b_i(\delta)$ are given in \eqref{eq:SimEquaAiBi}. 
From \eqref{eq:EqA} we can find $\psi_e$:
\begin{gather}
-\sin\delta \left(a_1(\delta)\sin(\psi)+b_1(\delta)\cos(\psi)\right)=0\\
\rightarrow \psi_e=\arctan\left(-\frac{b_1(\delta_e)}{a_1(\delta_e)}\right)\label{eq:HePsi}
\end{gather}
Substituting $\psi_e(\delta_e)$ into equation \eqref{eq:EqB}, one can solve for  $\beta_e$:
\begin{equation}
    \beta_e=\arcsin\left(-\omega b_2(\delta_e)\cos(\psi_e)-a_2(\delta_e)\sin(\psi_e)\right)\label{eq:HeBeta}
\end{equation}
\textcolor{black}{This gives up to two solutions $\beta_e$ for each pair of $\delta_e,\psi_e$
.}\\
From \eqref{eq:EqD}, one can solve for $\theta_e$:
\begin{equation}
    \theta_e=\arccos\left({\frac{\alpha \delta_e}{\omega a_4(\delta_e)}}\right)\label{eq:HeTheta}
\end{equation}
\textcolor{black}{We obtain a single solution of $\theta_e$ in the range $[0,\pi)$ for a given  $\delta_e$.} 
Substituting the solutions of $ \psi,  \beta,\theta$ from \eqref{eq:HePsi},\eqref{eq:HeBeta},\eqref{eq:HeTheta} back into equation  \eqref{eq:EqC} gives a scalar equation in $\delta_e,\omega$:
\begin{equation}
\begin{array}{l}
\frac{\alpha \delta}{a_4} \left( a_3 - \left( \frac{s_\psi b_1}{\sqrt{a_1^2 + b_1^2}} a_2 - \frac{s_\psi a_1}{\sqrt{a_1^2 + b_1^2}} b_2 \right) \frac{s_\psi a_1}{\sqrt{a_1^2 + b_1^2}} \right) \\[12pt]
= s_\beta \sqrt{1 - \omega^2 \left( \frac{s_\psi b_1}{\sqrt{a_1^2 + b_1^2}} a_2 - \frac{s_\psi a_1}{\sqrt{a_1^2 + b_1^2}} b_2 \right)^2} \frac{-s_\psi b_1}{\sqrt{a_1^2 + b_1^2}} \end{array}
\label{eq:scalatHe}
\end{equation}
where $a_i,b_i$ are functions of $\delta$ from \eqref{eq:SimEquaAiBi} and $s_\psi=\pm 1$ describes the two solutions of $\psi$ from eq. \eqref{eq:HePsi}, such that $\psi_-=\psi_+-\pi$, and $s_\beta=\mathrm{sign}(\cos \beta_e)$ describes the two solutions of $\psi$ from eq. \eqref{eq:HeBeta}. Eq. \eqref{eq:scalatHe} gives a scalar equation which is transcendental in $\delta_e$, and can be solved only numerically for given $\omega$. On the other hand, Eq. \eqref{eq:scalatHe} can be solved analytically for $\omega^2$ under given $\delta_e$.

{\bf Solution multiplicity and symmetries:}\\
Solving equation \eqref{eq:HePsi} gives two possible solutions for $\psi_e$, represented by $s_\psi=\pm1$. Nevertheless, after choosing a specific solution for $\psi_e$, only one solution for $\beta_e$ exists in \eqref{eq:HeBeta}. This is because substituting back into \eqref{eq:EqC}, it can be shown that $\cos \beta_e$ must have a specific sign. 
For a given frequency $\omega$, equation \eqref{eq:scalatHe} is transcendental in $\delta$, and may give multiple solutions. It can be shown that these solutions are independent of the choice of $s_\psi=\pm1$ for solution multiplicity of $\psi_e,\beta_e$ in \eqref{eq:HePsi} ,\eqref{eq:HeBeta}. Within the range $0<\delta<\pi$ of the joint angle, we show below that there are up to two solutions for \eqref{eq:scalatHe} under given $\omega$, which are denoted by $\delta_1,\delta_2$. For each of these solutions, there is a single solution of \eqref{eq:HeTheta} for $\theta_e$, and two pairs of solutions of \eqref{eq:HePsi} and \eqref{eq:HeBeta} for $(\beta_e,\psi_e)$, depending on the sign $s_\psi=\pm1$. Thus, upon varying $\omega$, we obtain four branches of solutions, denoted by $\{1+\}$, $\{1-\}$, $\{2+\}$ and $\{2-\}$.

{\bf Solutions existence:} 
Equations \eqref{eq:HeBeta},\eqref{eq:HeTheta} and \eqref{eq:scalatHe} give three inequalities which determine the range of existence of the solution:
\begin{gather}
  \left|-\omega b_2(\delta_e)\cos(\psi_e)-a_2(\delta_e)\sin(\psi_e)\right|\le1  \label{eq:spatialExist1}\\
  \left|\frac{\alpha \delta_e}{\omega a_4(\delta_e)}\right|\le1
  \label{eq:spatialExist2}\\
  \left(\frac{a_3 }{a_4}\scalemath{0.6}{+}\frac{a_1(a_1 b_2 -a_2 b_1)}{a_4(a_1^2+b_1^2)}\right)^2\frac{\alpha^2 \delta^2}{b_1^2} \scalemath{0.6}{-}a_1^2\scalemath{0.6}{-}b_1^2 \ge 0
  \label{eq:spatialExist3}
\end{gather}

{\bf Stability analysis:} 
We now examine stability of the spatial helical solutions using linear analysis, by the Jacobian matrix, calculated analytically. 
\begin{gather}
    \Bo{J}=\left(\frac{d [\Bo{A}^{-1}\Bo{g}]}{d \Bo{q}}\right)_{\Bo{q}=\Bo{q_e}}
    \label{eq:jacobian}
\end{gather}
In this case, $\Bo{J}$ is a full matrix, rather than block-diagonal as in \eqref{eq:TmJac}. Figure \ref{fig:DeltaStab} shows the solution branches of $\delta_e$ as function of the nondimensional frequency $\omega$, for both planar tumbling and spatial helical motions. 
The {planar tumbling} solution curve is in blue color, and the {spatial helical} solutions are in red color, two solutions denoted as $\delta_1$ (stable branch, solid curve) and $\delta_2$ (unstable, dashed). 
It can be seen that the {spatial helical} solution starts in the same frequency that the {planar tumbling} solution loses its stability, at the bifurcation point where  $\omega=\omega_0$. Another bifurcation occurs where  the unstable part of the planar tumbling branch bifurcates into an unstable branch $\delta_2$ of spatial helical motion.  Those unstable solutions are practically irrelevant, since they are not observed in experiments or attained in numerical simulations. The stable branch $\delta_1$ of spatial helical solution exists in a region from $\omega_0$ to $\omega_s$, which is the Step-out frequency, that can be found from the existence conditions \eqref{eq:spatialExist1},~\eqref{eq:spatialExist2}. 
In frequencies higher than the step-out frequency $\omega_s$, we obtain an asynchronous solution, for which the angles $\Bo{q}_a(t)$ converge to a steady-state periodic solution, rather than constant values. This means that the swimmer's rotation loses synchronization with that of the magnetic field, so that the phase difference $\beta(t)$ is oscillating periodically rather than converging to a constant value. Similarly, the joint angle $\delta(t)$ is also oscillating periodically rather than converging to a constant value. This asynchronous solution is represented by the simulation results in Figure \ref{fig:Asin}, and the rightmost region in Figure \ref{fig:Speed}. This solution
is also similar to the asynchronous solution obtained in \cite{morozov2014chiral} for a rigid magnetically-chiral cylinder. 
Analyzing the linearization eigenvalues of $\Bo{J}$ in \eqref{eq:jacobian}, it can be shown that upon crossing $\omega=\omega_s$, a complex conjugate pair of eigenvalues turns into purely imaginary eigenvalues. This indicates that a Hopf bifurcation point of the synchronous solution occurs at the step-out frequency $\omega=\omega_s$ \cite{Thesis}.  Using the physical parameter values in \eqref{eq:paraArt}, the nondimensional step-out frequency $\omega_s$ corresponds to physical frequency value of $\Omega_s=210 [Hz]$. This value was beyond the range of applicable frequencies with the experimental setup in \cite{wu2021helical}, up to $50 [Hz]$.\\

\begin{figure}
	\centering
	\includegraphics[width=0.7\linewidth]{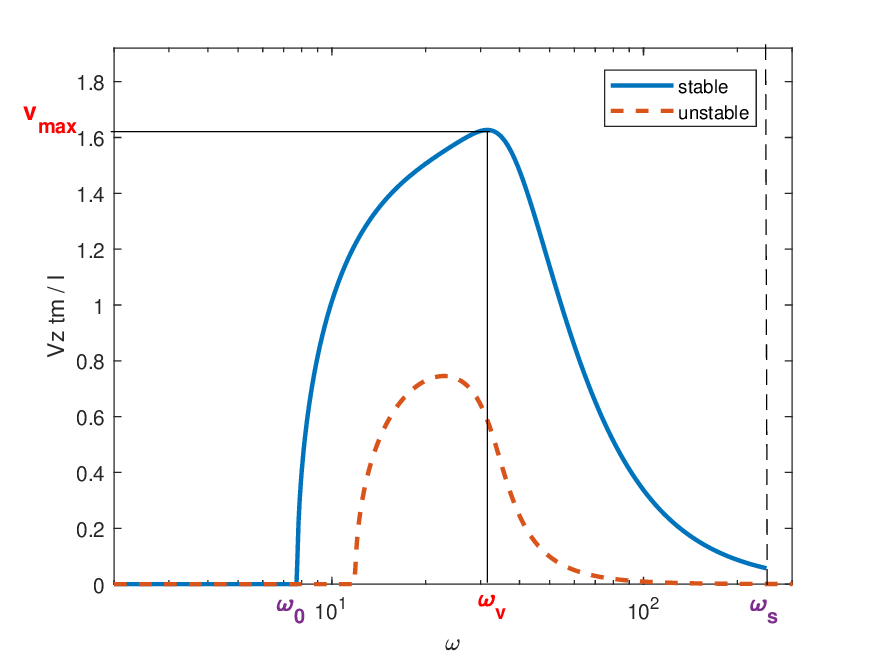}
	\caption{Normalized Speed $\frac{t_m}{l}V_z$ of the synchronous solution as a function of nondimensional frequency  $\omega$.}
	\label{fig:Vz}
\end{figure}

\begin{figure}
	\centering
	\includegraphics[width=0.7\linewidth]{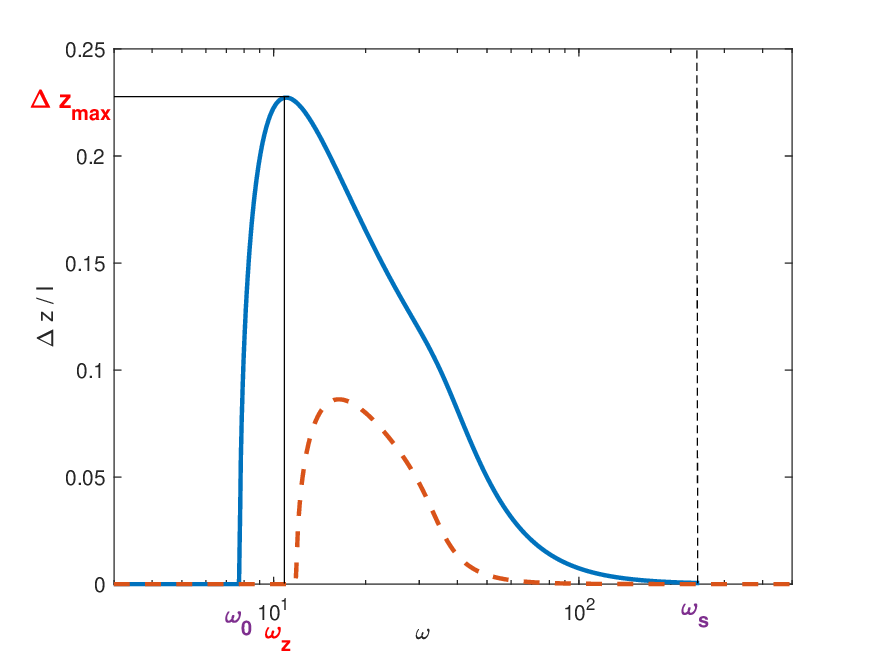}
	\caption{Normalized pitch $\Delta z/ l$ of the synchronous solution as a function of nondimensional frequency  $\omega$.}
		\label{fig:Dz}
\end{figure}

{\bf Swimming speed:} 
We now present expression of the swimming speed $V_z$. 
Applying the simplifying assumptions on the velocity equations \eqref{eq:Velocity} gives an expression for the forward speed, in $\Bo{\hat{z}}$ direction:
\begin{equation}
    V_z=\frac{\omega \sin(2\theta_e)(\cos(\delta_e)+1)(\cos(\delta_e+\psi_e)-\cos(\psi_e))}{8c_{n1}l_1(3-\cos(\delta_e))}
    \label{eq:speed}
\end{equation}
One can also examine the pitch, the displacement per cycle, defined as  $\Delta z=V_z\frac{2\pi}{\omega}$.\\
The {planar tumbling} solution with $\sin \theta_e =0$,  gives zero propulsion $V_z=0$. By substituting the {spatial helical} solution into  \eqref{eq:speed}, one obtains an expression for $V_z$ that only depends on the joint angle $\delta_e$, which, in turn, depends on the nondimensional frequency $\omega$. Note that solution multiplicity $s_\psi=\pm1$ for $\psi_e$ simply results in reversing the sign of $V_z$, so that solutions $\{i-\}$ are identical to solutions $\{i+\}$, except for swimming in $-\Bo{\hat{z}}$ direction rather than $+\Bo{\hat{z}}$. 
Figures \ref{fig:Vz},\ref{fig:Dz}, show the speed $V_z$ and pitch $\Delta z$ as a function of $\omega$, for both stable (solid) and unstable (dashed) branches of spatial helical motion (for $\omega<\omega_0$, the only solution is planar tumbling so that $V_z=0$). It can be seen that for the stable branch, one obtains maximal values of $V_z$ and $\Delta z$ for two different optimal frequencies, denoted by $\omega_v,\omega_z$.  Using the physical parameter values in \eqref{eq:paraArt}, the nondimensional values of maximal swimming speed and pitch and their corresponding optimal frequencies from Figures \ref{fig:Vz} and \ref{fig:Dz} are obtained as: 
\begin{equation} \begin{array}{l}
     V_{max}=21.2 [\mu m / sec],\; \Omega_v=26.8 [Hz]  \\[8pt]
     \Delta z_{max}=1.63  [\mu m/ cycle], \; \Omega_z=9.27 [Hz].
\end{array}
\label{eq:optimal_V_Z}    
\end{equation}
The values of $V_{max}$ and $\Omega_v$ are reasonably comparable to the estimated values from the nano-swimmer's experimental measurements in \cite{wu2021helical} ($V_{max} \approx 28 [Hz]$, and $\Omega_v \approx 28 [\mu m / sec]$, respectively). 
In higher frequencies, above the step-out frequency $\omega_s$, the {spatial helical} solution ceases to exists and an asynchronous solution begins. The asynchronous solution's mean speed and pitch continue the line smoothly from the synchronous {spatial helical} solution \cite{Thesis}. Furthermore, in \cite{Thesis} we also developed an approximate formulation of the helical motion using perturbation expansion at the limit of low stiffness, allowing us to obtain explicit approximate expressions for optimal frequency and swimming speed.

\begin{figure}
	\centering
	\includegraphics[width=0.7\linewidth]{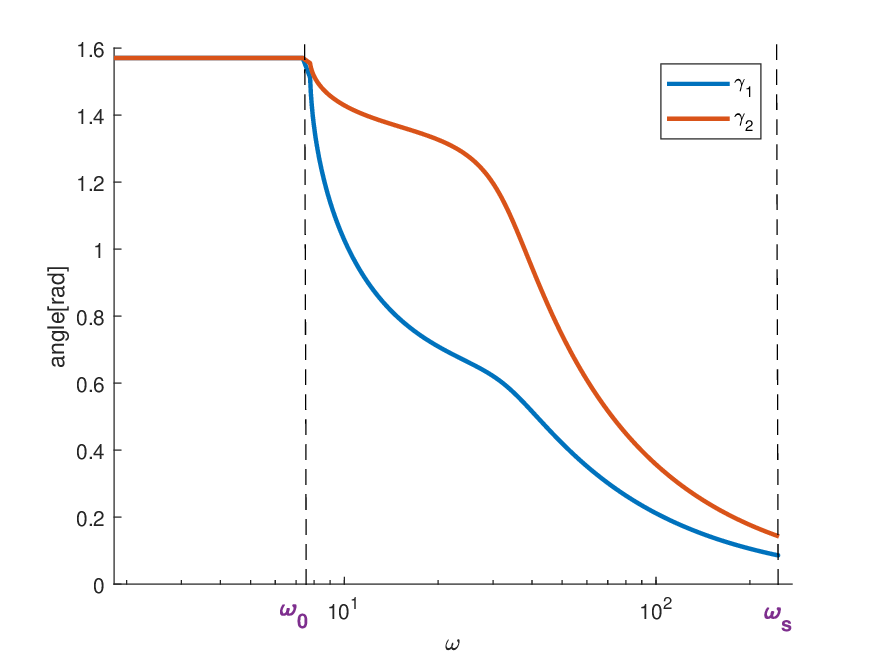}
	\caption{ Cone angels $\gamma_1,\gamma_2$ of the stable synchronous solution as a function of nondimensional frequency  $\omega$.}
		\label{fig:Cone}
\end{figure}

{\bf Cone angles:} At steady-state synchronous rotational motion of the nano-swimmer, the constant joint angle $\delta=\delta_e$, cannot be directly measured from the video capture of 2D images of the experiments. On the other hand, the two links are rotating along a cone envelope centered about $\Bo{\hat z}$ axis, with constant angles $\gamma_1,\gamma_2$, as depicted inside Figure \ref{fig:DeltaStab}. These two angles can be estimated from the experiments' video movies. Using the nano-swimmer's kinematics in \eqref{eq:Rot1}, the expressions for these angles can be obtained as:
\begin{equation}
    \begin{array}{ll}
         \gamma_1=\cos^{-1}[\sin \theta_e \cdot \sin \psi_e]  \\[8pt]
         \gamma_2=\cos^{-1}[\sin \theta_e \cdot \sin (\psi_e+\delta_e)] . 
    \end{array}
\end{equation}
Figure \ref{fig:Cone} plots the two cone angles $\gamma_i$ as a function of the nondimensional frequency $\omega$ under the stable branches of the planar tumbling and spatial helical solutions. In the regime of planar tumbling, $0<\omega<\omega_0$, the two cone angles are $\gamma_i=90^\circ$, since the nano-swimmer lies within $\hat{x}-\hat{y}$ plane. In the regime of spatial helical motion $\omega_0<\omega<\omega_s$, both angles decrease gradually with increasing $\omega$, until losing synchrony at $\omega=\omega_s$. This behavior has also been observed from the experiments in \cite{wu2021helical}. 

\begin{figure*}     
\centering
\subfigure{}{\includegraphics[width=0.49\linewidth]{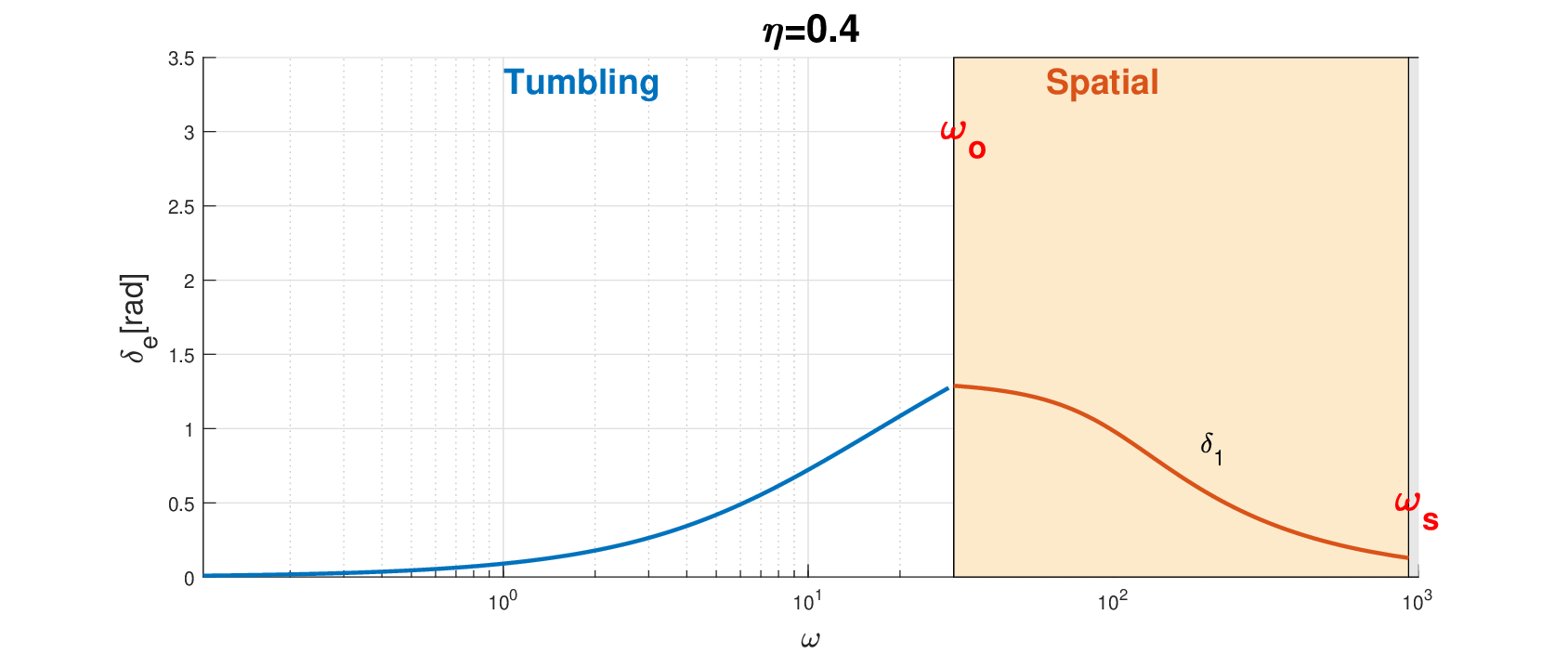}}
\subfigure{}{\includegraphics[width=0.49\linewidth]{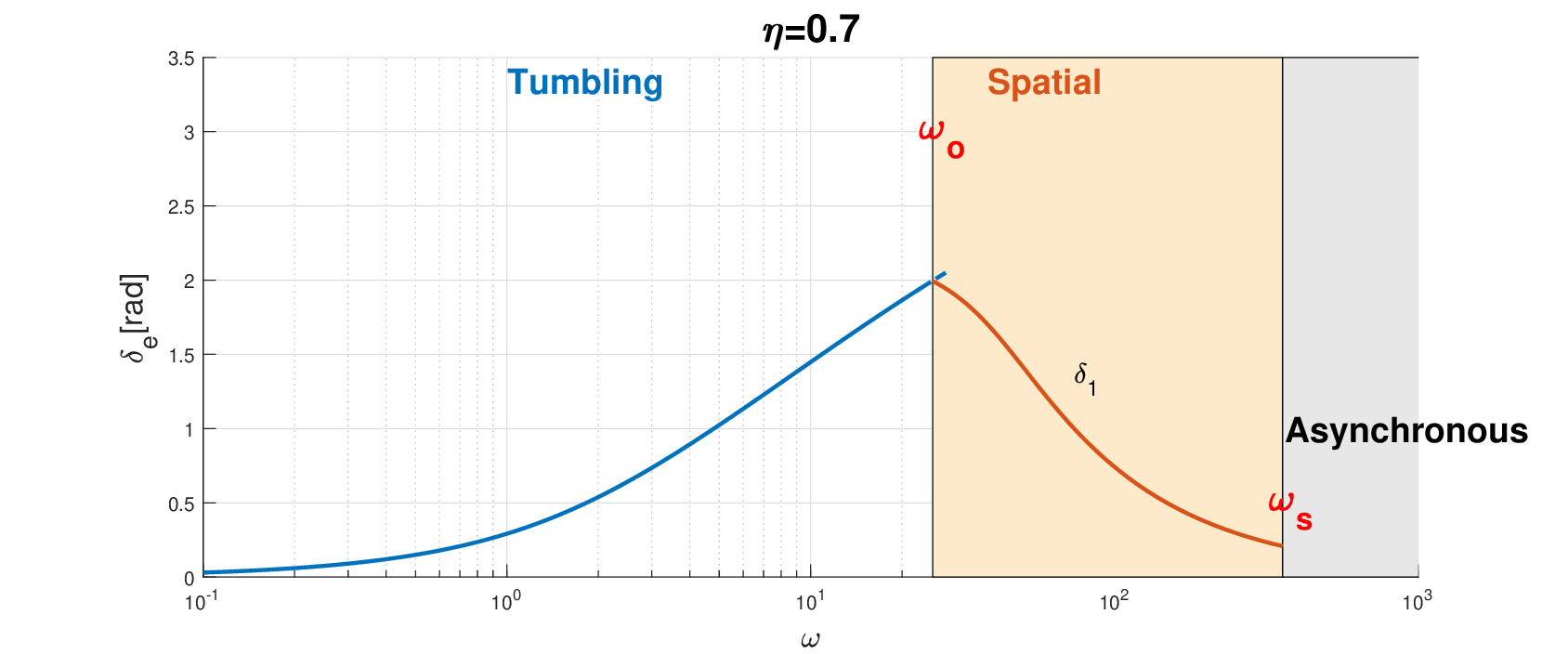}}
\subfigure{}{\includegraphics[width=0.49\linewidth]{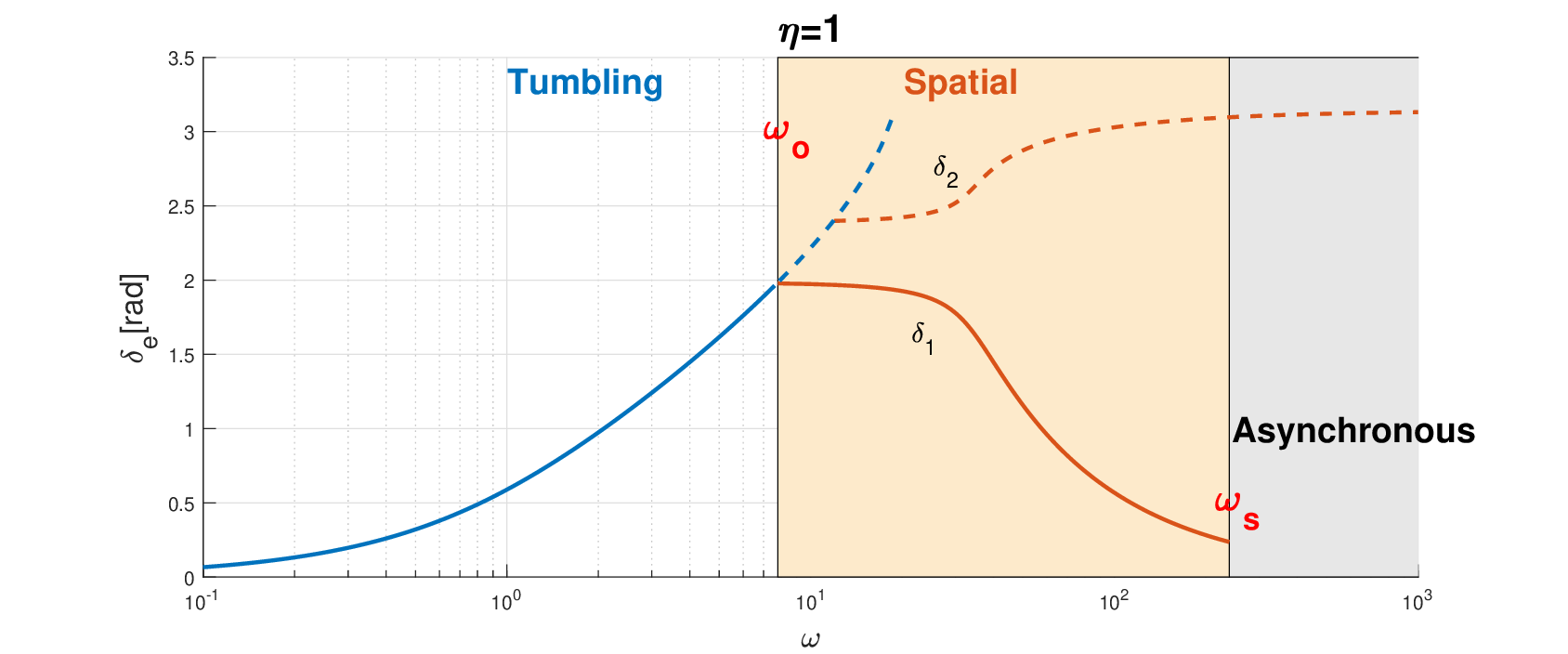}}
\subfigure{}{\includegraphics[width=0.49\linewidth]{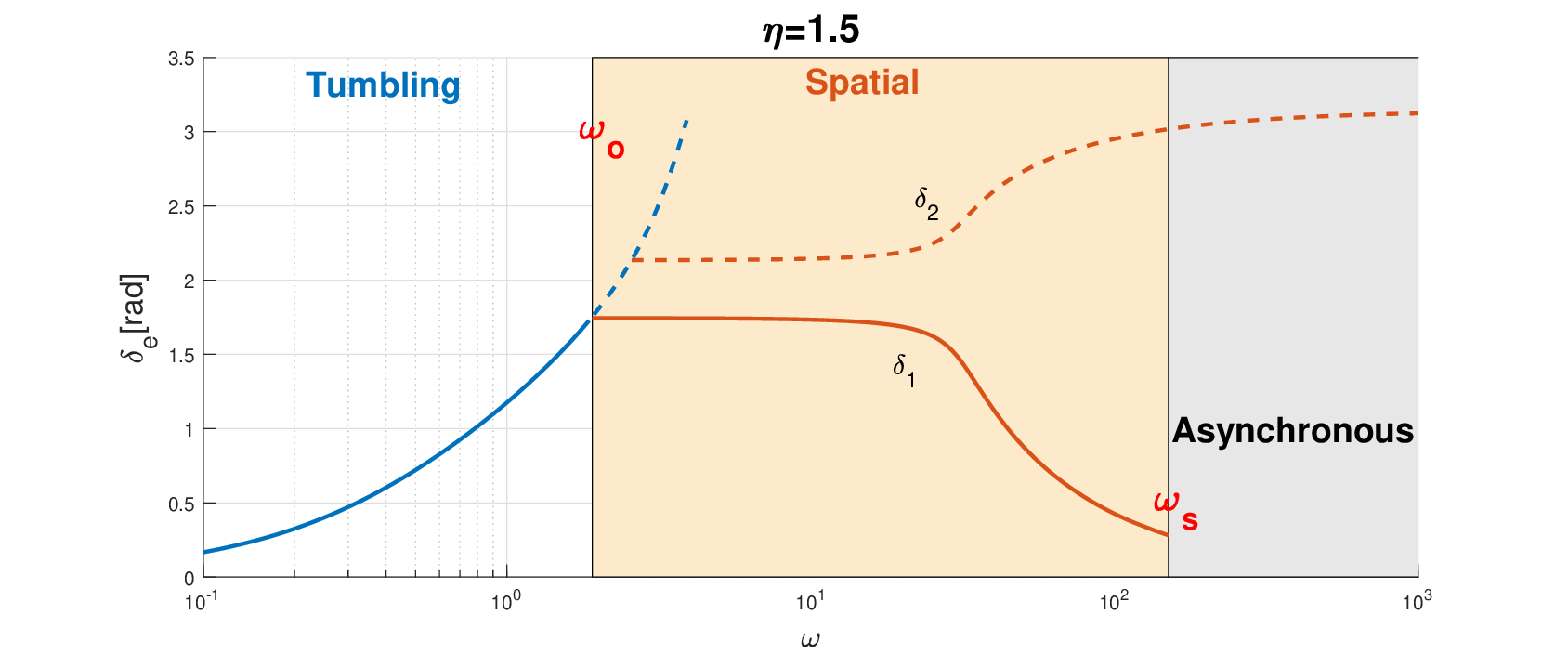}}
\subfigure{}{\includegraphics[width=0.49\linewidth]{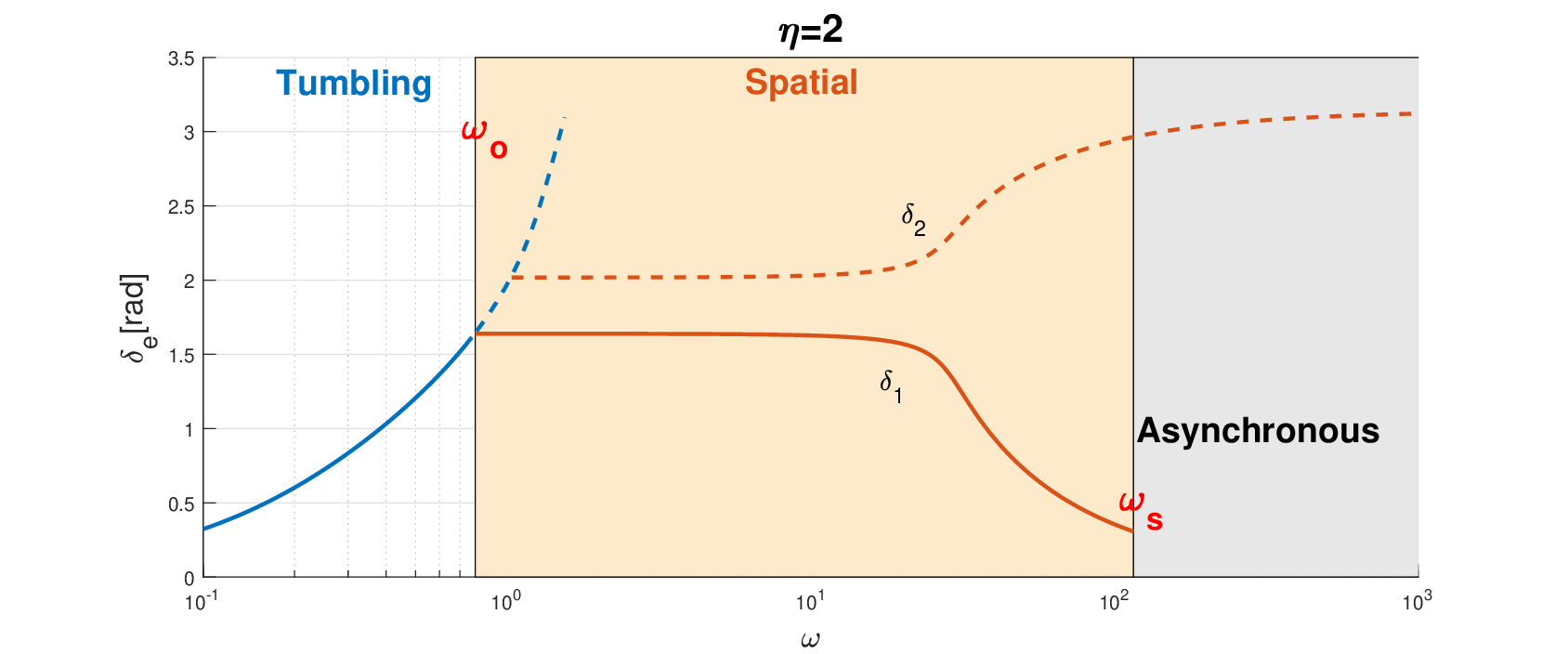}}
\subfigure{}{\includegraphics[width=0.49\linewidth]{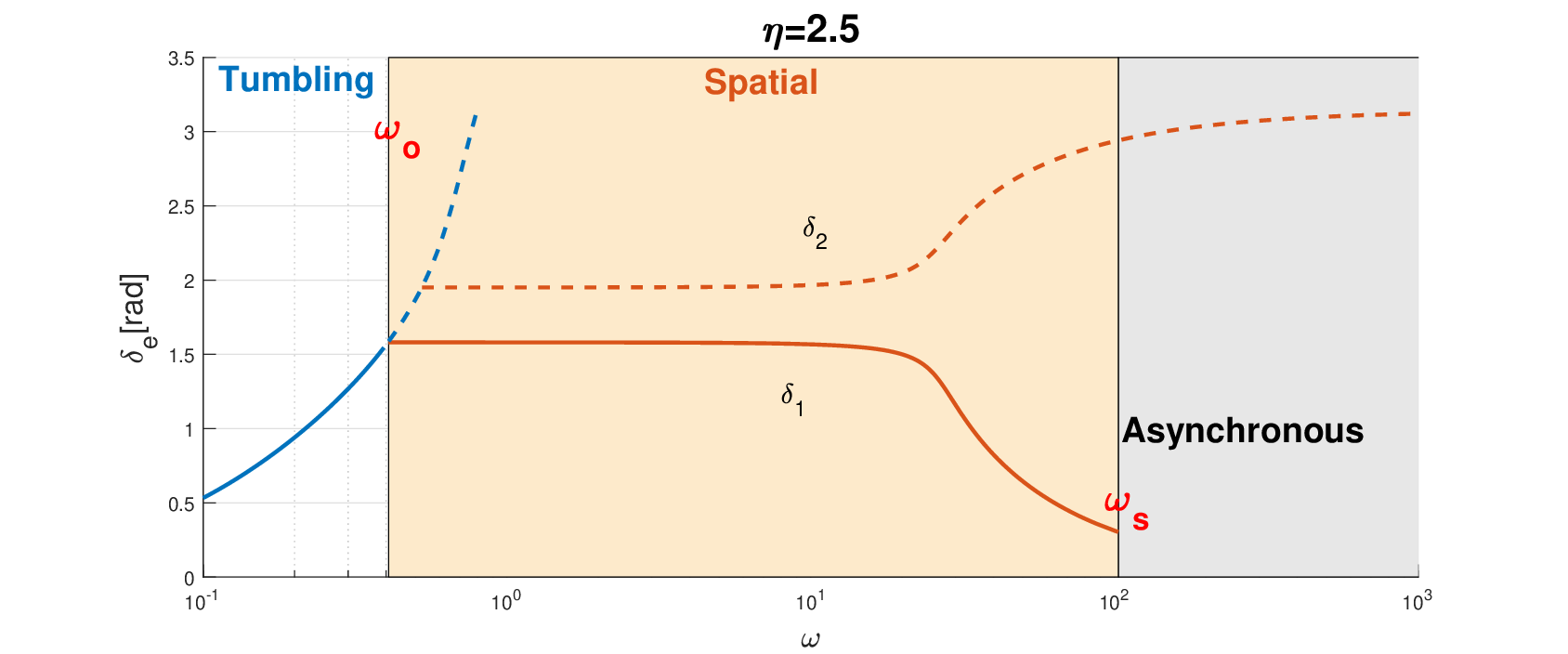}}
    \caption{Synchronous solutions of the joint angle $\delta_e$ as a function of the nondimensional frequency $\omega$, for different values of links' length ratio $\eta$. Blue curve - planar tumbling; red curves - branches of spatial helical solution. Solid curves - stable solutions; dashed curves - unstable solutions.}
    \label{fig:SolEta}
\end{figure*}
\section{Analysis of generalized cases} \label{sec.general_case}
In this section, we study generalized cases where some of the simplifying assumptions from previous section are relaxed. First, we analyze the cases of unequal link’s length and of a conically rotating magnetic field. Next, we show some numerical optimizations for those and other cases. 
\subsection{Unequal links' length}
\begin{figure}[t]
    \centering
    \includegraphics[width=0.7\linewidth]{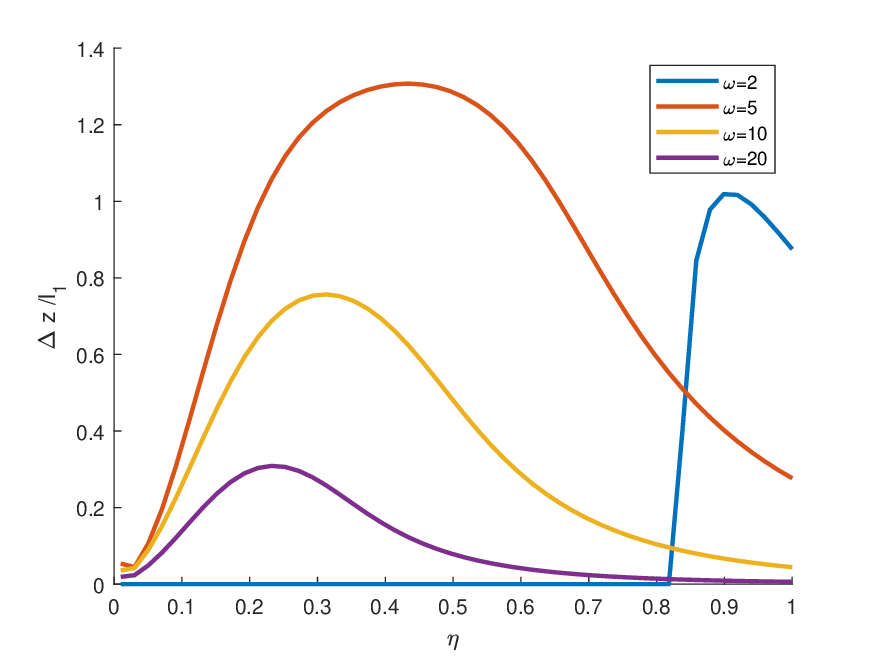}
    \caption{Normalized pitch $\Delta z/l_1$ in the stable solution branch, plotted as function of links’ length ratio $\eta$ under different fixed values of nondimensional frequency $\omega$. }
    \label{fig:5.1}
\end{figure}

\begin{figure}
    \includegraphics[width=0.7\linewidth]{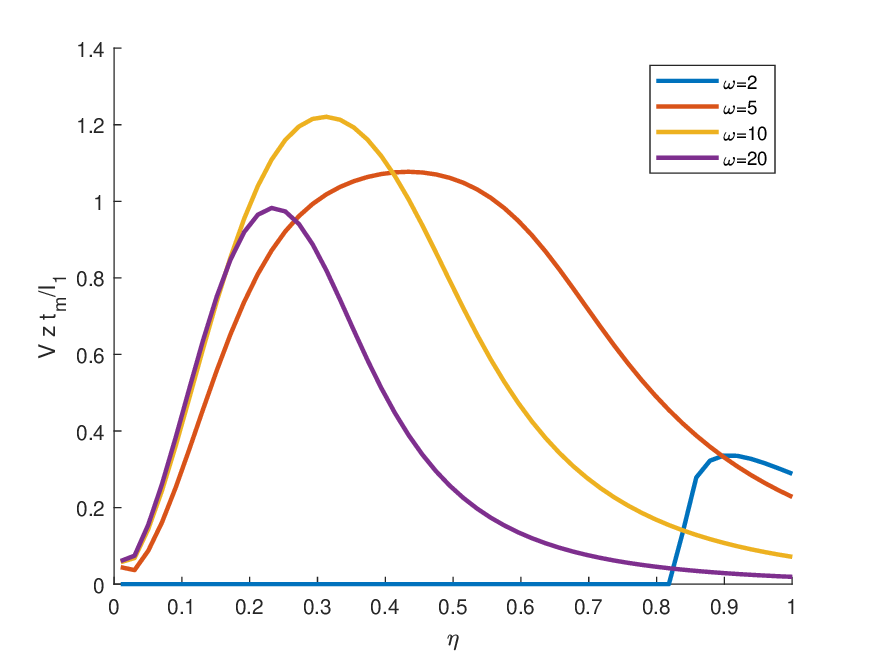}
    \caption{Normalized Speed $\frac{t_m}{l_1}V_z$ in the stable solution branch, plotted as function of links’ length ratio $\eta$ under different fixed values of nondimensional frequency $\omega$.}
    \label{fig:5.2}
\end{figure}

We now study the effect of unequal links’ lengths. We define a new variable for the ratio of the lengths $\eta=l_2/l_1 $. We revisit the equilibrium equations \eqref{eq:SimpEQAqg} applying the assumptions from section \ref{Sec:SA} above, while assuming $l_2=\eta l_1$. Note that this also implies scaling on the drag coefficients, as:
\begin{equation}
    c_{n_2}=\eta c_{n_1};c_{t_i}=\half c_{n_i};d_{n_i}=\frac{1}{12}c_{n_i}l_i^2;d_{t_i}=0;\MagConic=0 \label{eq:FuLe1}
\end{equation}
From \eqref{eq:FuLe1}, we obtain the generalized equilibrium equations. Those equations are the same as in \eqref{eq:4DOFEQ}. However, the expressions for $a_i(\delta_e),b_i(\delta_e)$ are different from \eqref{eq:SimEquaAiBi}, and are given in Table \ref{eq:EtaEquationAiBi} below. 
\begin{table}[b]
\caption{\label{eq:EtaEquationAiBi}%
}
\begin{ruledtabular}
\begin{tabular}{ccccc}
$a_1(\delta_e,\eta)=-\frac{\eta \sin \delta_e (\eta+4)}{12(\eta +1)}$\\
$b_1(\delta_e,\eta)=\frac{ \sin \delta_e (6+\eta\cos\delta_e(7+\eta))}{24(\eta +1)}$\\
$a_2(\delta_e,\eta)=\frac{\eta^2 \sin \delta_e (3+\eta\cos\delta_e(4+\eta))}{12(\eta +1)}$\\
$b_2(\delta_e,\eta)=-\frac{\eta^3 \cos^2 \delta_e (4+\eta)+ 6\eta^2\cos\delta_e+1+4\eta}{12(\eta +1)}$\\
$a_3(\delta_e,\eta)=-\frac{(\eta+1)\left(2+6\eta+2\eta^2+6\eta^3+2\eta^4+12\eta^2\cos\delta_e(2-2\eta+2\eta^2)\right)}{20\eta+8\eta^2-4\eta\cos^2\delta_e+8}$\\
$a_4(\delta_e,\eta)=-\frac{\eta^2\left(8\eta+\eta^3+8\eta^2+cos\delta_e(6+6\eta+2\eta\cos\delta_e)\right)}{4\left(5\eta+2\eta^2-\eta\cos^2\delta_e+2\right)}$
\end{tabular}
\end{ruledtabular}
\end{table}
In fact, the expressions in Table \ref{eq:EtaEquationAiBi} are generalization of \eqref{eq:SimEquaAiBi} to the case of $\eta \ne 1$. \\
We are able to solve the equilibrium equations \eqref{eq:4DOFEQ} using precisely the same steps taken above for $\eta=1$. Nevertheless, the multiplicity of solution branches and their existence conditions now depend also on the ratio $\eta$.\\
In Figure \ref{fig:SolEta}, we present the branches of planar tumbling and spatial helical solutions for the joint angle $\delta=\delta_e$ as a function of the nondimensional frequency $\omega$, under different values of links' length ratio $\eta$. Solid lines denote stable solutions while dashed lines denote unstable ones. From the plots, we can see that in low values of $\eta<1$, i.e. shorter tail link, the planar tumbling solution exists in a wider range of nondimensional frequencies, whereas the spatial helical solution emerges at a larger $\omega$. In addition, for low $\eta$ only the stable solution branches exist, and the joint angle $\delta_e$ attains smaller values. In larger values of $\eta$, unstable solution branches emerge and both $\omega_0$ and $\omega_s$ are decreasing with growing $\eta$. We can also notice that the maximal joint angle $\delta_e$ is increasing with $\eta$ up to $\eta\approx0.8,\delta_e\approx2[rad]$. After this point, $\delta_e$ decreases until reaching $\delta_e \to \pi/2$ for the stable solution.\\
Next, we examine the influence of $\eta$ on swimming speed. The expression for $V_z$ remains the same as in \eqref{eq:speed}, except that the equilibrium angles $\delta_e,\theta_e,\psi_e$ now depend also on $\eta$. 
In Figures \ref{fig:5.1} and \ref{fig:5.2}, we plot the normalized pitch $\Delta z/l_1$ and speed $\frac{t_m}{l_1}V_z$ as a function of the links' length ratio $\eta$ for given fixed frequencies. We obtain maximal pitch or speed, which are attained at different values of $\eta$ depending on frequency $\omega$. The plots indicate that a combined optimum exists in both the frequency $\omega$ and  length ratio $\eta$, which is further examined below in Figures \ref{fig:etaPitch} and \ref{fig:etaSpeed}.

\subsection{Conically rotating magnetic field}
We now consider the case of conically rotating magnetic field with a nonzero component $h$ in \eqref{eq:MagField}, while assuming equal links' lengths $\eta=1$. 
The equilibrium equations \eqref{eq:4DOFEQ} are generalized as:
\begin{gather}
    \omega \sin\theta_e \sin\delta_e \left(a_1(\delta_e)\sin(\psi_e)+b_1(\delta_e)\cos(\psi_e)\right)=0\nonumber\\
    \sin\theta_e\left(\sin\beta_e +\omega b_2(\delta_e)\cos(\psi_e)+a_2(\delta_e)\sin(\psi_e)\right)\nonumber\\ \textcolor{blue}{+\MagConic \cos\theta_e}=0\nonumber\\
    \cos\beta_e\sin\psi_e +\cos\theta_e\left(\sin\beta_e\cos\psi_e+\omega a_3(\delta_e)\right)\nonumber\\\textcolor{blue}{-\MagConic \cos\psi_e\sin\theta_e}=0\nonumber\\
    \alpha\delta_e=\omega \cos\theta_e a_4(\delta_e)
    \label{eq:ConicEqations}
\end{gather}
While $a_i(\delta_e),b_i(\delta_e)$ remain the same as in \eqref{eq:SimEquaAiBi}.
The equilibrium equations \eqref{eq:ConicEqations} are different from the previous cases, and the added effect of the constant component $h$ is marked in \textcolor{blue}{blue}. That is, when setting $h=0$, equations \eqref{eq:ConicEqations} reduce back to \eqref{eq:EqA}-\eqref{eq:EqD}.
Equations \eqref{eq:ConicEqations} are solved in a similar way as in the case of $h=0$ above. Importantly, it can be shown that $\sin\theta_e=0$ is no longer a solution of \eqref{eq:ConicEqations}, in contrast to the previous case with $ \MagConic=0$. This implies that the planar tumbling solution no longer exists for $h\ne0$. 
The solution of first equation of \eqref{eq:ConicEqations} for spatial helical motion remains:
\begin{equation}
    \psi_e=\arctan\left(-\frac{b_1(\delta_e)}{a_1(\delta_e)}\right)\label{eq:CocPsie}
\end{equation}
The solution of the $4^{th}$ equation of \eqref{eq:ConicEqations} also remains:
\begin{equation}
    \theta_e=\arccos\left(\frac{\alpha \delta_e}{\omega a_4(\delta_e)}\right)\label{eq:CocThetae}
\end{equation}
From the second equation in \eqref{eq:ConicEqations} we obtain the new solution for $\beta_e$:
\begin{equation}
    \beta_e=\arcsin\left( \text{-}\omega(b_2(\delta_e)\cos(\psi_e)\text{+}a_2(\delta_e)\sin(\psi_e))\textcolor{blue}{-\frac{\MagConic \cos\theta_e}{\sin\theta_e}}\right)\label{eq:CocBetae}
\end{equation}
Finally, by substituting the solutions \eqref{eq:CocPsie}, \eqref{eq:CocThetae} and \eqref{eq:CocBetae} into the third equation of \eqref{eq:ConicEqations}, we obtain a scalar equation in $\delta,\omega$. This equation is transcendental in $\delta_e$ and high-degree polynomial in $\omega$, which can be solved numerically for known $\delta_e$ or $\omega$.
For given nondimensional frequency $\omega$, we numerically obtain up to two solutions for $\delta_e$ within the range $(0,\pi)$. Importantly, the addition of $h \ne 0$ violates previous symmetries of solutions with respect to the choice of $s_\psi=\pm 1$ for multiplicity of the solutions for $\psi_e$ and $\beta_e$. That is, we obtain four different solution branches $\{1+\}$, $\{1-\}$, $\{2+\}$ and $\{2-\}$, which do not satisfy symmetry interrelations as in the case of $h=0$. Stability of the solution branches can be determined by calculating eigenvalues of the Jacobian matrix, as defined in \eqref{eq:jacobian}.

\begin{figure}
	\centering
	\includegraphics[width=1\linewidth]{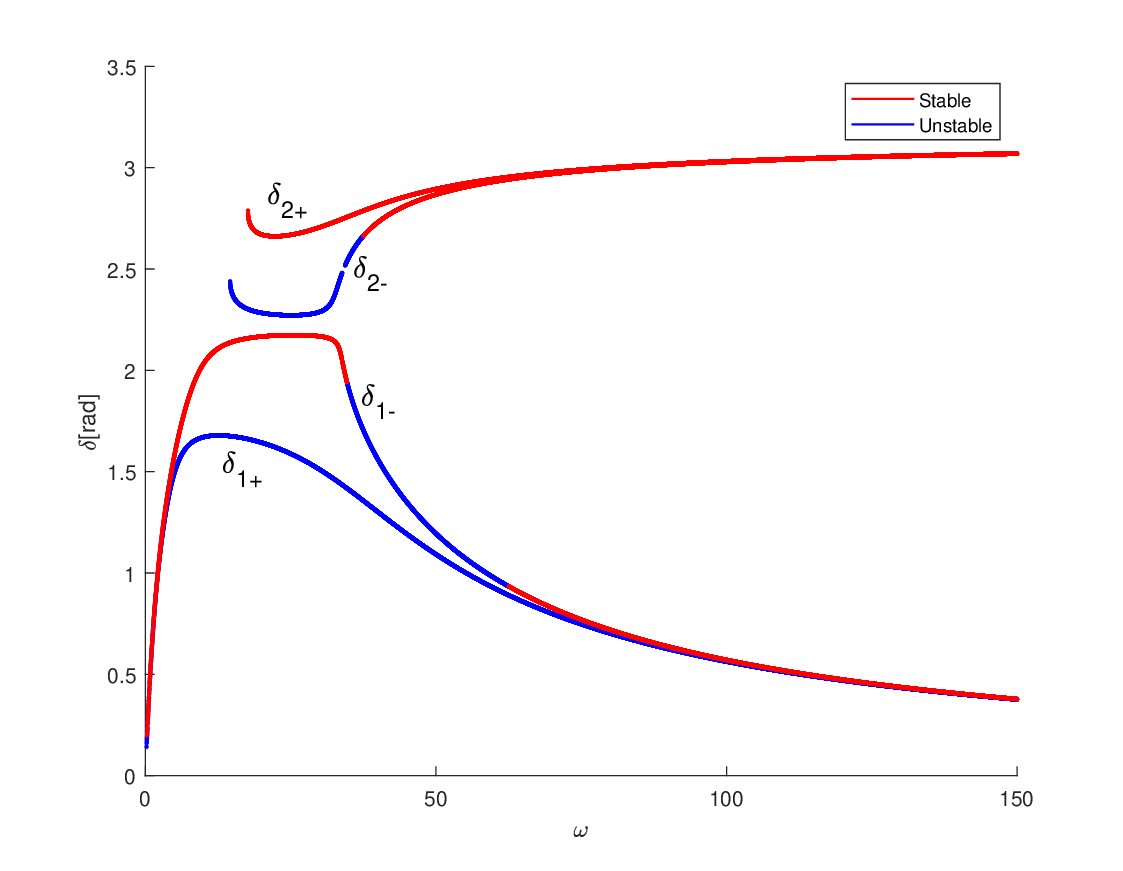}
	\caption{Solution branches for the joint angle $\delta_e$ as a function of nondimensional frequency  $\omega$, under a conically-rotating magnetic field with constant component $h=0.2$. Blue curves - stable solution. Red curves - unstable solution.}
	\label{fig:DeltaCOnic02}
\end{figure}

\begin{figure}
	\centering
		\includegraphics[width=1\linewidth]{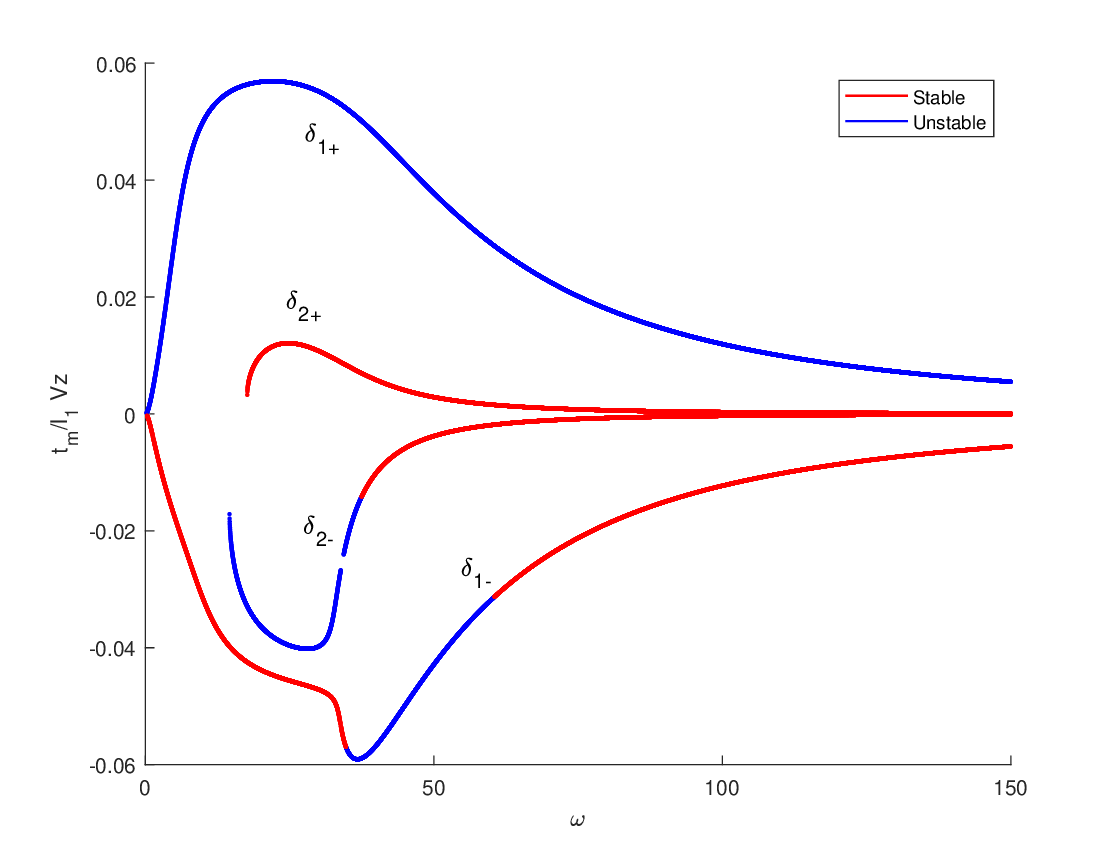}
	\caption{Normalized swimming speed $\frac{t_m}{l} V_z$ for all solution branches as a function of nondimensional frequency  $\omega$, under a conically-rotating magnetic field with constant component $h=0.2$. Blue curves - stable solution. Red curves - unstable solution.}
		\label{fig:DZ01}
\end{figure}

Figure \ref{fig:DeltaCOnic02} plots the four different solution branches of the joint angle  $\delta_e$ as a function of the nondimensional frequency  $\omega$, under a conically-rotating magnetic field with constant component $h=0.2$. Stable solutions are marked as blue curves while unstable solutions are marked as red curves. It can be seen that solution branch $\{1+\}$ is stable and solution branch $\{2+\}$ is unstable for the entire range of nondimensional frequencies $\omega$, similarly to the previous case of $h=0$. However, branches $\{1-\}$ and $\{2-\}$ include stability transitions upon varying $\omega$. This interesting phenomenon is similar to stability transitions of the same micro-swimmer model under a \emph{planar} oscillating magnetic field, which were studied in \cite{paul2023nonlinear}. Figure \ref{fig:DZ01} plots the normalized swimming speed $\frac{t_m}{l}V_z$ as a function of the nondimensional frequency for all solution branches, under the same case of $h=0.2$. It can be seen that the solution branches  $\{i+\}$ have $V_z>0$ and branches  $\{i-\}$ have $V_z<0$. Each branch has an optimal frequency for which $|V_z|$ is maximized, but symmetry between $\pm$ branches is lost. 

\begin{figure}
	\centering
	\includegraphics[width=1\linewidth]{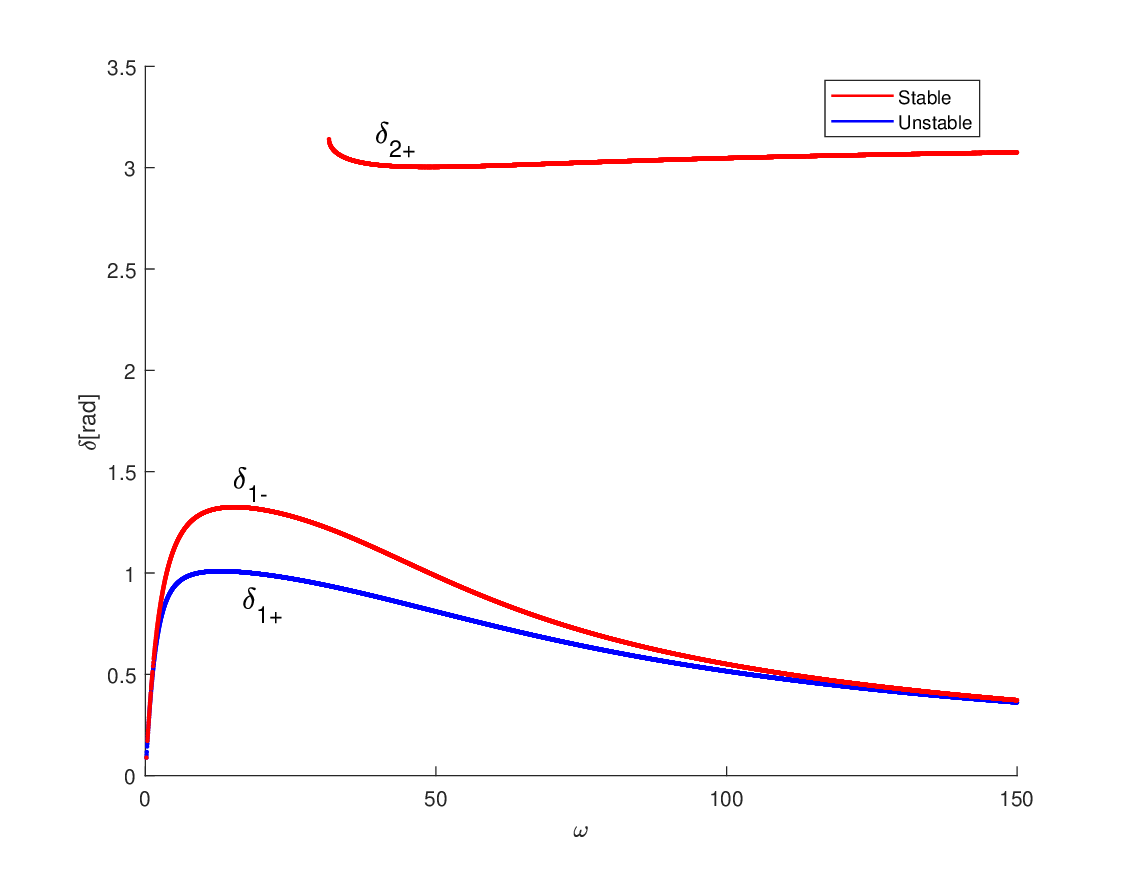}
	\caption{Solution branches for the joint angle $\delta_e$ as a function of nondimensional frequency  $\omega$, under a conically-rotating magnetic field with constant component $h=1$. Blue curves - stable solution. Red curves - unstable solution.}
	\label{fig:DeltaCOnic1}
\end{figure}

\begin{figure}
			\includegraphics[width=1\linewidth]{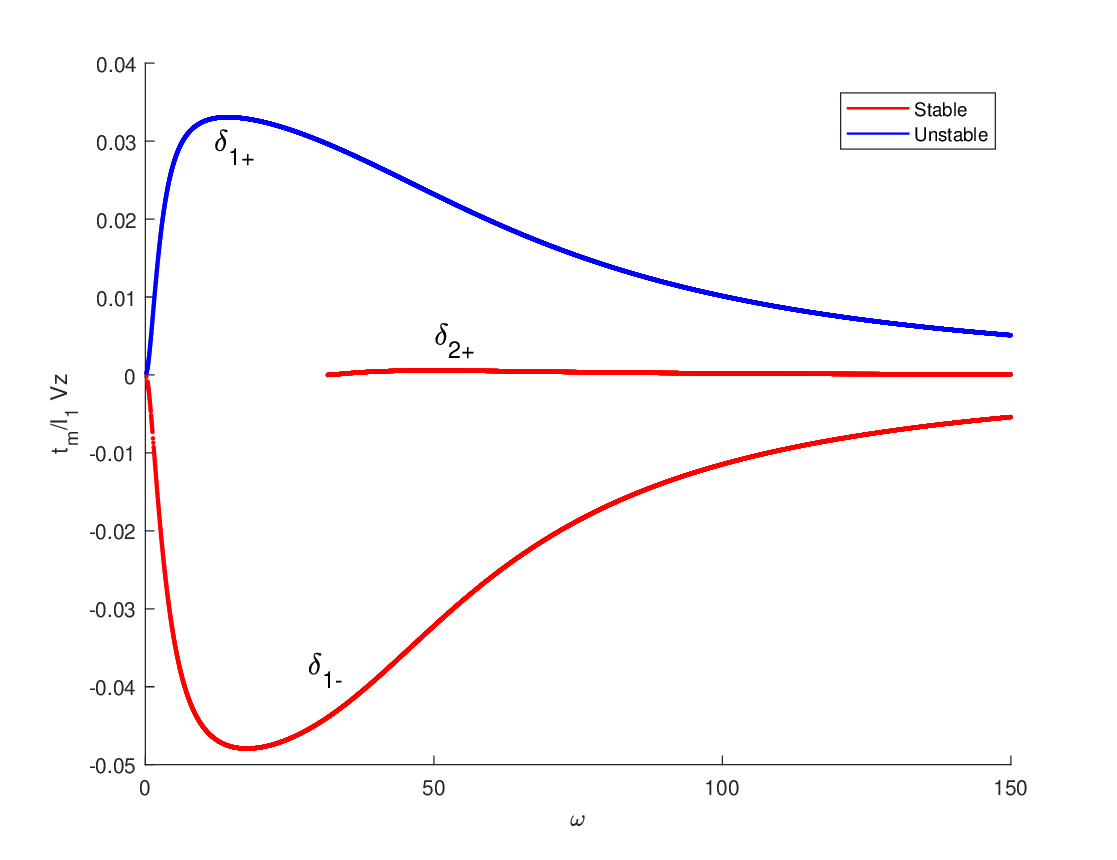}
	\caption{Normalized swimming speed $\frac{t_m}{l} V_z$ for all solution branches as a function of nondimensional frequency  $\omega$, under a conically-rotating magnetic field with constant component $h=1$. Blue curves - stable solution. Red curves - unstable solution.}
	\label{fig:DZ00}
\end{figure}

Figure \ref{fig:DeltaCOnic1} plots all solution branches of the joint angle  $\delta_e$ as a function of the nondimensional frequency  $\omega$ for a 
larger constant component, $h=1$. The normalized swimming speed $\frac{t_m}{l}V_z$ as a function of the nondimensional frequency in all solution branches for $h=1$ is plotted in Figure \ref{fig:DZ00}. Stable solutions are marked as blue curves while unstable solutions are marked as red curves. It can be seen that the solution branch $\{2-\}$ no longer exists for this case, and that the maximal swimming speed for this case is larger than that of $h=0.2$. This motivates combined optimization of swimming speed upon varying  both frequnecy and $h$, as discussed below.


\subsection{Numerical parametric optimizations}
We now demonstrate optimization of the nano-swimmer's performance with respect to various physical parameters, combined with actuation frequency. In each case, we set nominal values of physical parameters as given in \eqref{eq:paraArt}, except for one chosen parameter which is varied, along with the actuation frequency $\Omega$. We obtain the swimmer's steady-state solution of spatial helical motion, calculate the mean speed $V_z$ and pitch $\Delta z$ and show both of them in contour plots. We consider only solution branch $\{1+\}$, which is stable for all $\Omega$. (In cases where $h=0$, the solution branch $\{1-\}$ is a mirror reflection of $\{1+\}$, with same speed $|V_z|$ in negative direction). 

\begin{figure}
    \includegraphics[width=1\linewidth]{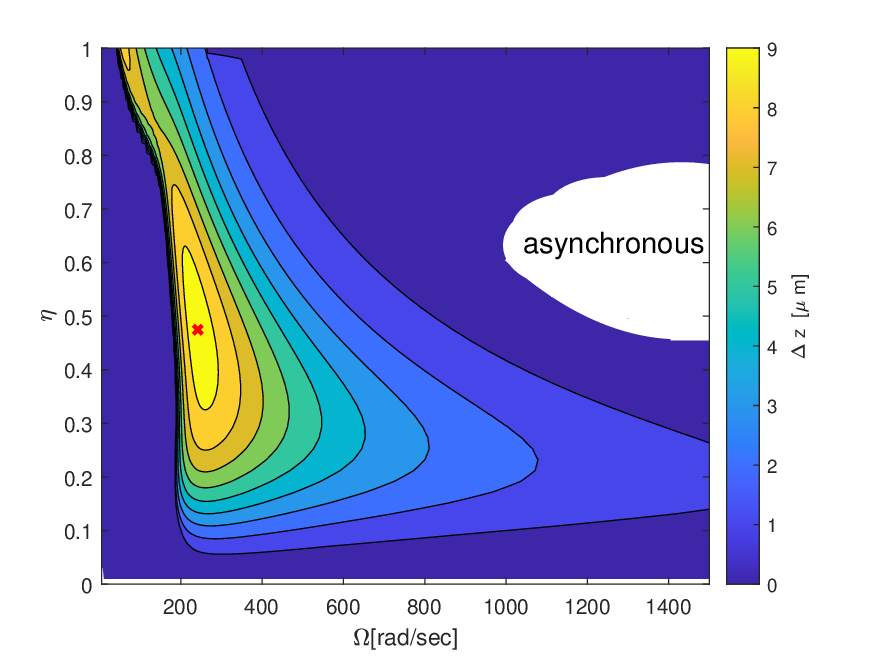}
    \caption{Contour map of the pitch $\Delta z$ as function of the frequency $\Omega$ and links’ length ratio $\eta$ }
    \label{fig:etaPitch}
\end{figure}

\begin{figure}
    \includegraphics[width=1\linewidth]{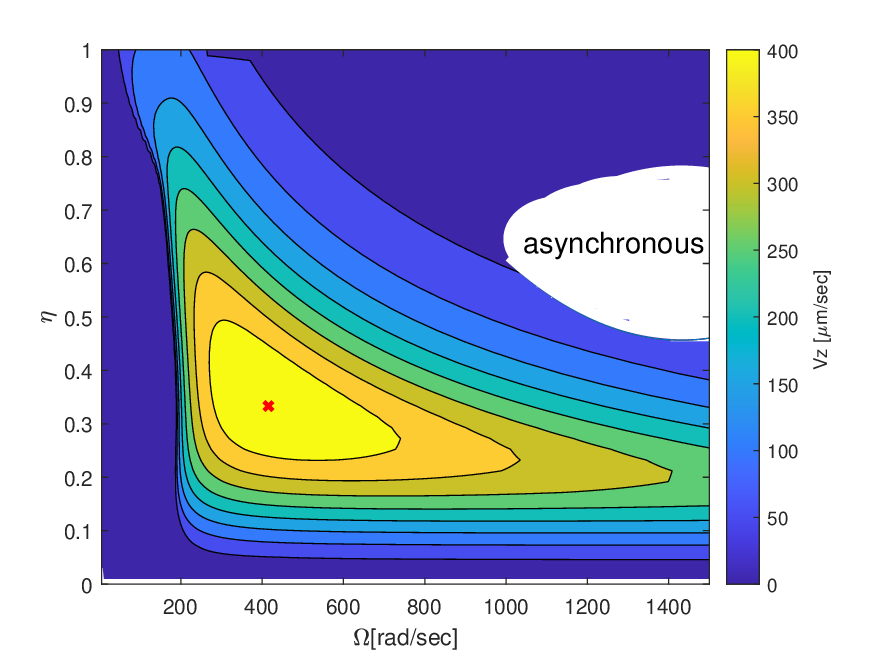}
    \caption{Contour map of the speed $V_z$ as function of the frequency $\Omega$ and links’ length ratio $\eta$}
    \label{fig:etaSpeed}
\end{figure}

First, we vary the links’ length ratio $\eta$ while the head link's length $l_1$ is held constant. Figures \ref{fig:etaPitch} and \ref{fig:etaSpeed} show contour maps for the pitch $\Delta z$ and speed $V_z$ as a function of the frequency $\Omega$ and links' length ratio $\eta$.
In Figure \ref{fig:etaPitch}, there is an optimum point for the pitch  $\Delta z_{max}=9.48 [\mu m/ cycle]$, at $\eta=0.45,\Omega=241.7 [rad/sec]=38.46 [Hz]$. This is improvement of $\sim 5.8$ times compared to the nominal case in \eqref{eq:paraArt}, \eqref{eq:optimal_V_Z}. 
In Figure \ref{fig:etaSpeed}, there is an optimum point for the speed  $\Delta V_{max}=448.6 [\mu m/ sec]$, at $\eta=0.33,\Omega=415.5 [rad/sec]=66.1 [Hz]$. This is improvement of $\sim~21$ times compared to the nominal case in \eqref{eq:paraArt}, \eqref{eq:optimal_V_Z}, although it is attained at relatively large frequency, beyond those applied experimentally in \cite{wu2021helical}. The white region in both figures represents the asynchronous regime beyond step-out frequency. 

\begin{figure}
    \centering
    \includegraphics[width=1\linewidth]{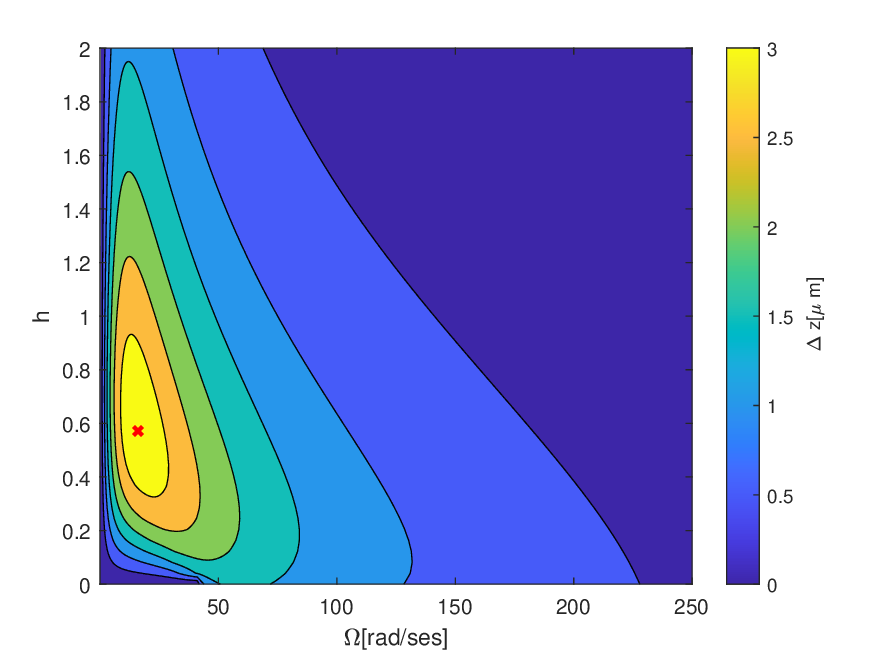}
    \caption{Contour map of the pitch $\Delta z$ as function of the frequency $\Omega$ and conic field component $h$.}
    \label{fig:conicPitch}
\end{figure}

\begin{figure}
    \centering
    \includegraphics[width=1\linewidth]{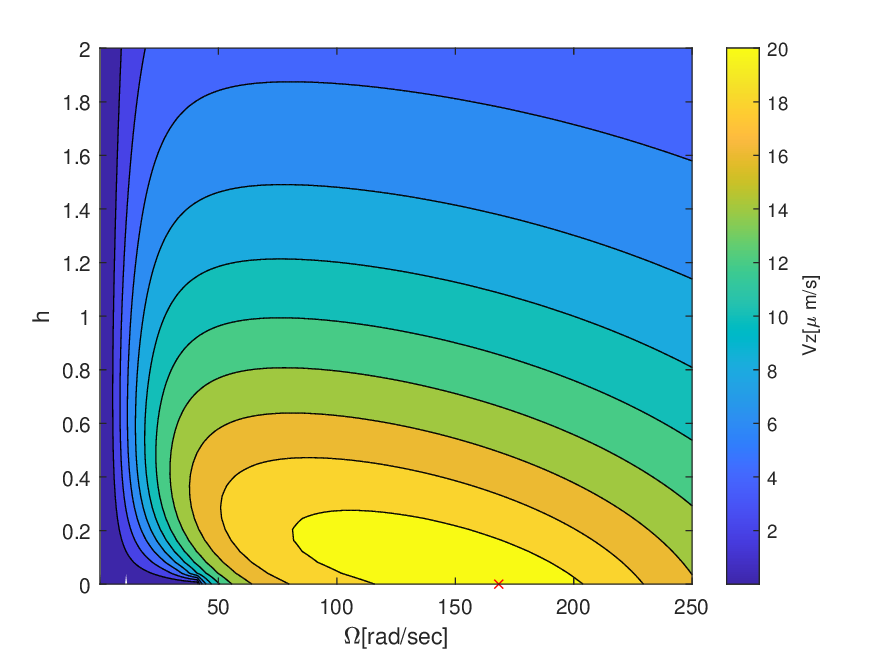}
    \caption{Contour map of the speed $V_z$ as function of the frequency $\Omega$ and conic field component $h$.}
    \label{fig:conicSpeed}
\end{figure}

Next, we examine the case of conically-rotating magnetic field, with varying constant component $h$. 
Contour plots of the pitch $\Delta z$ and speed $V_z$ as a function of frequency $\Omega$ and $h$ are shown in Figures \ref{fig:conicPitch} and \ref{fig:conicSpeed}.
In Figure \ref{fig:conicPitch}, there is an optimum for the pitch  $\Delta z_{max}=3.36$,  at $h=0.57,\; \Omega=16[rad/sec]=2.5]Hz]$. This is improvement of $~2$ times compared to the nominal case in \eqref{eq:paraArt}, \eqref{eq:optimal_V_Z}. (Note that a similar optimum has been found in \cite{mirzae2020modeling} using multi-bead numerical simulations).
 However, Figure \ref{fig:conicSpeed} shows that the optimum for the speed $V_z$ is obtained at $h=0$, at the same values given in \eqref{eq:optimal_V_Z}. 

\begin{figure}
    \centering
    \includegraphics[width=1\linewidth]{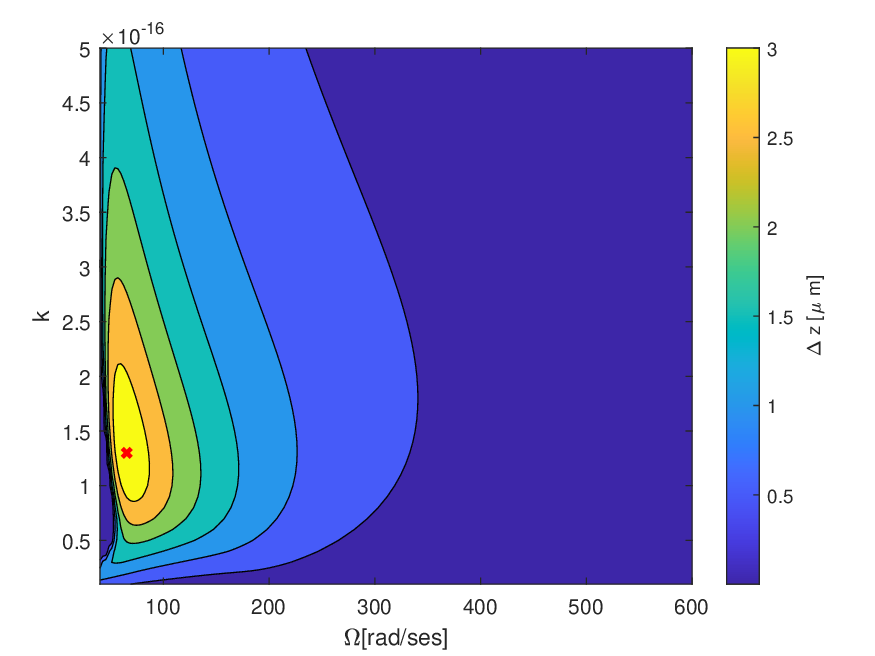}
    \caption{Contour map of the pitch $\Delta z$ as function of the frequency $\Omega$ and joint's stiffness component $k$}
    \label{fig:KPitch}
\end{figure}

\begin{figure}
    \centering
    \includegraphics[width=1\linewidth]{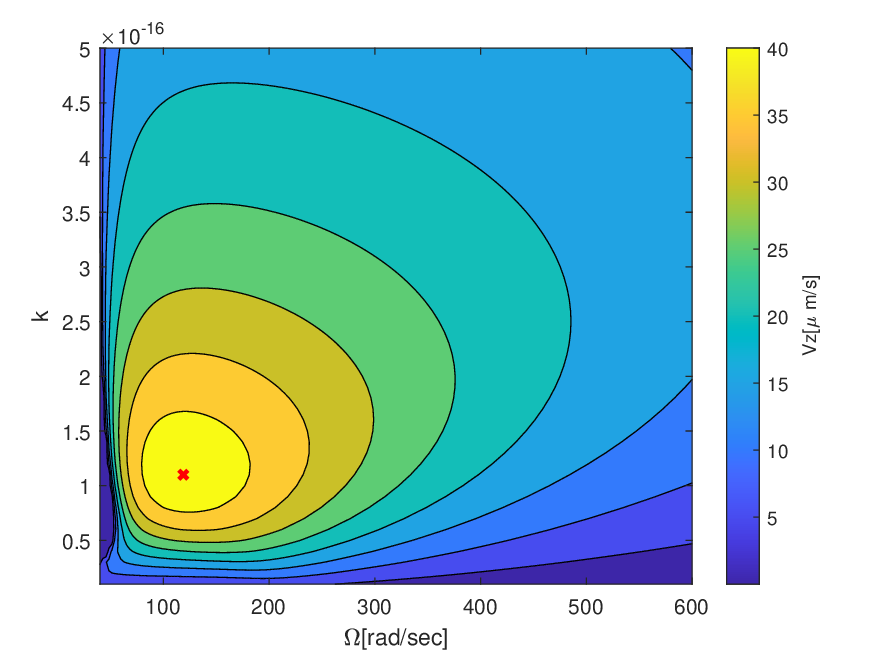}
    \caption{Contour map of the speed $V_z$ as function of the frequency $\Omega$ and joint's stiffness component $k$}
    \label{fig:KSpeed}
\end{figure}

Next, we examine optimization with respect to the joint's torsional stiffness $k$. Contour plots of $\Delta z$ and $V_z$ as a function of $\Omega$ and $k$ are shown in Figures \ref{fig:KPitch} and \ref{fig:KSpeed}.
In Figure \ref{fig:KPitch}, there is an optimum for the pitch  $\Delta z_{max}=3.36 [\mu m/ cycle]$,  at $k=1.3\cdot 10^{-16}[N\cdot m],\;\Omega=65.3[rad/sec]=10.4[Hz]$. This is improvement of $\sim 2$ times compared to the nominal case in \eqref{eq:paraArt}, \eqref{eq:optimal_V_Z}. 
In Figure \ref{fig:KSpeed}, there is an optimum for the speed  $V_{max}=43.46 [\mu m/ cycle]$, at $k=1.1\cdot 10^{-16}[N\cdot m],\; \Omega=119[rad/sec]=19[Hz]$. This is improvement of $\sim 2$ times compared to the nominal case in \eqref{eq:paraArt}, \eqref{eq:optimal_V_Z}.

\begin{figure}
    \centering
    \includegraphics[width=1\linewidth]{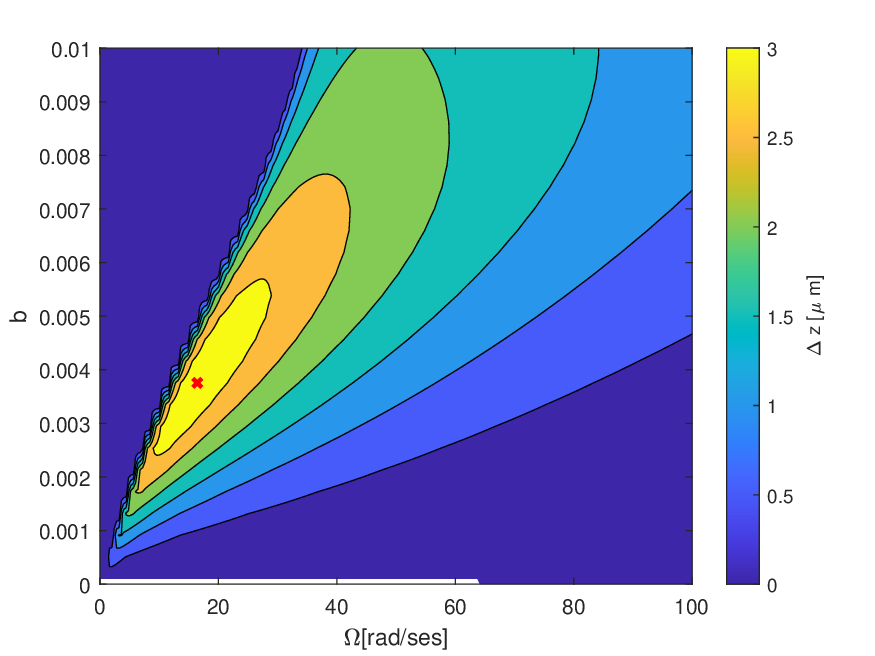}
    \caption{Contour map of the pitch $\Delta z$ as function of the frequency $\Omega$ and magnetic field's amplitude component $b$}
    \label{fig:BPitch}
\end{figure}

\begin{figure}[!t]
    \centering
    \includegraphics[width=1\linewidth]{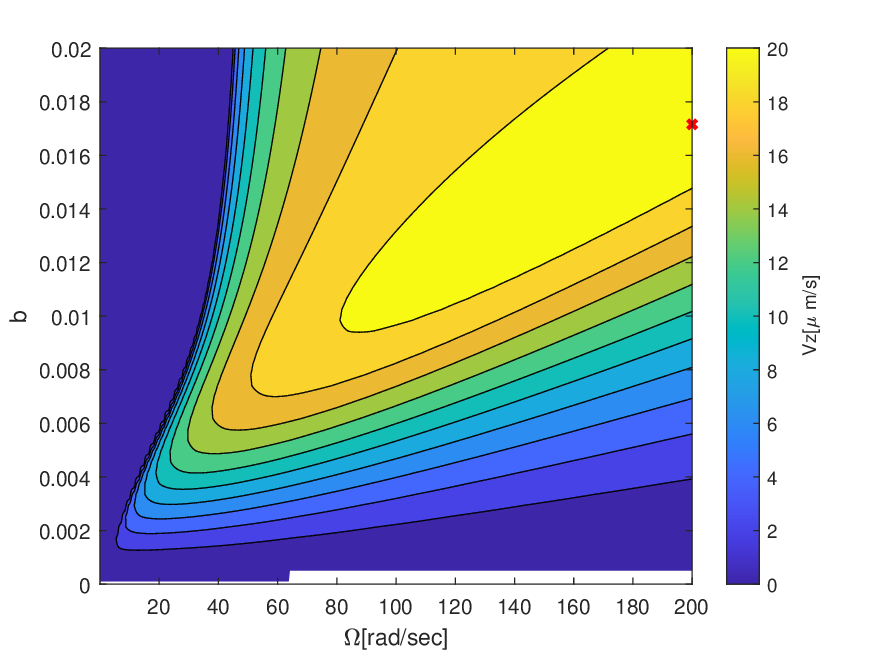}
    \caption{Contour map of the speed $V_z$ as function of the frequency $\Omega$ and magnetic field's amplitude component $b$}
    \label{fig:BSpeed}
\end{figure}

Finally, we examine optimization with respect to the magnetic field's magnitude  $b$. 
Contour plots of $\Delta z$ and $V_z$ as a function of $\Omega$ and $b$ are shown in Figures \ref{fig:BPitch} and \ref{fig:BSpeed}.
In Figure \ref{fig:BPitch}, there is an optimum for the pitch  $\Delta z_{max}=3.36 [\mu m/ cycle]$,  at $b=3.7[mT],\Omega=16[rad/sec]$. This is improvement of $\sim 2$ times compared to the nominal case in \eqref{eq:paraArt}, \eqref{eq:optimal_V_Z}.
In Figure \ref{fig:BSpeed}, there is an optimum for the speed 
 $V_{max}=21.33 [\mu m/ cycle]$,
$V_z$ at $b=17.1[mT],\; \Omega=200[rad/sec]=31.4[Hz]$, at the limit of the plot. This is a very minor improvement compared to the nominal case in \eqref{eq:paraArt}, \eqref{eq:optimal_V_Z}, which was attained at $b=15[mT]$.  Some more numerical optimization results with respect to other physical parameters can be found in \cite{Thesis}.

\section{Concluding discussion}
In this work, we revisited the simplified two-link model of a nano-swimmer with flexible hinge under a rotating magnetic field, which was studied numerically and demonstrated experimentally in \cite{wu2021helical}. We explicitly formulated and analyzed the nano-swimmer's nonlinear dynamic equations of spatial motion. We reduced the dynamic equations to a 4-DOF time-invariant system using transformation of variables. We obtained explicit analytic solutions for synchronous motion under simplifying assumptions, for both regimes of in-plane tumbling and spatial helical corkscrew swimming. We conducted stability analysis of the solution branches, and found stability transitions and bifurcations of multiple solution branches.  We obtained an explicit expression for the forward speed and pitch, and found optimal frequencies maximizing speed and pitch.
Finally, we presented  analysis of the influence of additional effects such as unequal links length and conically-rotating magnetic field, as well as numerical parametric optimization.\\

We now briefly discuss the limitations of our current analysis. First, the suggested swimmer model is simple and does not represent the fabricated swimmer accurately. It assumed a pointed uni-axial hinge, while the actual  hinge is a flexible beam that undergoes both bending and torsion. In addition, our model  did not account for hydrodynamic interaction between the links, and did not consider the influence of boundaries of the fluid domains. All those effects can be implemented into a new swimmer model in possible future extension of the research. Other possible generalizations can be interesting for future research, such as the effect of varying the directionality of link's magnetization, nano-swimmer composed of multiple magnetic links, flexible links, and additional  optimized choices in the time-varying profile of the external magnetic field \cite{el2020optimal}. In addition, the work can be extended to coordinated control of a group of magnetically-driven micro-propellers \cite{buzhardt2019dynamics}.\\

Despite all its limitations, this work showed a simple minimal model that captures most of dominating phenomena observed in experiments on the fabricated nano-swimmer, allowing us to better understand the dynamics of magnetic nano-swimmers. With this improved understanding, new and optimized magnetic nano-swimmers may be fabricated, which eventually will allow the designing of artificial magnetic nano-bots for bio-medical purposes advancing towards the grand challenge of micro-nano-biomedical robotics.

{\bf Acknowledgment:} This work has been supported by Israel Science Foundation, under grant no. 1382/23.

\section*{Appendix - grand resistance matrix of two-link structure}
\setcounter{equation}{0}
\renewcommand{\theequation}{A.\arabic{equation}}

In this appendix, we derive the grand resistance matrix of two rigidly-attached axisymmetric links, representing the nano-swimmer with immobilized joint. First, the grand resistance matrix of a rigid body with respect to a reference point $p$ relates hydrodynamic force and torque to the body's linear and angular velocities, as:
\begin{equation} \label{eq:grand_R}
    \left( \begin{array}{l}
         \Bo{F}  \\
         \Bo{M}_p 
    \end{array}\right) = 
    \underbrace{
    \left( \begin{array}{cc}
         \Bo{A} & \Bo{B}  \\
         \Bo{B}^T & \Bo{C} 
    \end{array}\right)}_{\Bo{Q}_p}
    \left( \begin{array}{l}
         \Bo{v}_p  \\
         \boldsymbol{\omega}
    \end{array}\right).
\end{equation}
We now explain how the grand resistance matrix is affected by rotation of reference frame and shift of the reference point $p$. Consider transformation of the vectors between two reference frames (1 and 2) using a rotation matrix $\Bo{R}$ such that $\Bo{F}^{(1)}=\Bo{RF}^{(2)}$, $\Bo{v}^{(1)}=\Bo{Rv}^{(2)}$, etc. It is straightforward to show that transformation of the grand resistance matrix $\Bo{Q}_p$ between the two reference frames is expressed as:
\begin{equation} \label{eq:trans_Q_R}
    \Bo{Q}_p^{(1)} = 
    \left( \begin{array}{cc}
         \Bo{R} & \Bo{0}  \\
         \Bo{0} & \Bo{R} 
    \end{array}\right) \Bo{Q}_p^{(2)} 
    \left( \begin{array}{cc}
         \Bo{R} & \Bo{0}  \\
         \Bo{0} & \Bo{R} 
    \end{array}\right)^T.
\end{equation}
Next, consider changing the reference point of a grand resistance matrix, between two points $c,p$ on the rigid body. The velocities of those two points are related as \begin{equation} \label{eq:trans_v}
    \Bo{v}_p=\Bo{v}_c+\boldsymbol{\omega} \times (\Bo{r}_p-\Bo{r}_c).
\end{equation} On the other hand, the total moments about $p$ and $c$ are related as \begin{equation} \label{eq:trans_M}
    \Bo{M}_p=\Bo{M}_c+ (\Bo{r}_c-\Bo{r}_p) \times \Bo{F}.
\end{equation}  Define the anti-symmetric matrix $\Bo{S}_{pc}$ such that $\Bo{S}_{pc}\Bo{v}=(\Bo{r}_p-\Bo{r}_c) \times \Bo{v}$ for all $\Bo{v}$. Using relations \eqref{eq:grand_R}, \eqref{eq:trans_v}  and \eqref{eq:trans_M}, the sub-blocks of the grand resistance matrices with two different reference points $\Bo{Q}_c$ and $\Bo{Q}_p$ are related as:
\begin{equation}
    \Bo{A}_p \!=\! \Bo{A}_c , \; \Bo{B}_p\!=\!\Bo{B}_c+\Bo{S}_{pc}^T\Bo{A}_c^T , \; \Bo{C}_p\!=\!\Bo{C}_c+\Bo{B}_c \Bo{S}_{pc} + \Bo{S}_{pc}^T \Bo{A}_c \Bo{S}_{pc} \;.
\end{equation}

We now calculate the grand resistance matrix of a ``rigidized'' two-link nano swimmer, with a relative angle $\delta$, as shown in Figure \ref{fig:model}. We assume two identical axisymmetric links of lengths $l$. In a reference frame located at the link's centers $c_i$ where $\Bo{\hat x}$ is aligned with the link's symmetry axis, the resistance matrices are diagonal, and their blocks are given as: 
\begin{equation}
\begin{array}{l}
\Bo{A}_{i,c_i}^{(i)} =-\left(
    \begin{array}{ccc}
      c_t   & 0 &0 \\
       0& c_n  & 0\\ 0 & 0 & c_n  
    \end{array} \right), \;
    \Bo{C}_{i,c_i}^{(i)} =-\left(
    \begin{array}{ccc}
      d_t   & 0 &0 \\
       0& d_n  & 0\\ 0 & 0 & d_n  
    \end{array} \right),\\[10pt]
       \mbox{and } \Bo{B}_{i,c_i}^{(i)}=0. \end{array}
\end{equation}
Note that the rotation-translation coupling matrices of each individual link $\Bo{B}_{i,c_i}$ vanish due to symmetry. We now wish to choose a common reference point at the joint $p$ connecting the two links. The relative displacement between the link's centers $c_i$ and $p$ expressed in the $i$-th link's frame are $\Bo{r}_{c_1}-\Bo{p}=(-b \; 0 \; 0)^T$ and  $\Bo{r}_{c_2}-\Bo{p}=(b \; 0 \; 0)^T$, where $b=l/2$. Using \eqref{eq:trans_v}, one can express the resistance matrices of both links with respect to the joint $p$. Next, we transform the resistance matrix of link 2 to the reference frame of link 1 using \eqref{eq:trans_Q_R}, where the rotation matrix between the links' frames is 
\begin{equation}
    \Bo{R}=
    \left(
    \begin{array}{ccc}
      \cos(\delta)   & \sin(\delta) &0 \\
       -\sin(\delta)& \cos(\delta)  & 0\\ 0 & 0 & 1  
    \end{array} \right).
\end{equation}
Finally, assuming negligible hydrodynamic interactions, the total resistance matrix of the rigidly-connected two links with respect to $p$ expressed in the frame of link 1 is simply obtained by summation, $\Bo{Q}_{tot,p}^{(1)}=\Bo{Q}_{1,p}^{(1)}+\Bo{Q}_{2,p}^{(1)}$. The blocks of this total resistance matrix are given in Table \ref{tab.R2}. 
\begin{table*} 
\centering
\[
\begin{array}{c}
\Bo{A}_{tot,p}^{(1)} =\left(
    \begin{array}{ccc}
      (1+\cos^2\delta)c_t + \sin^2\delta c_n    & (c_n-c_t)\cos\delta \sin \delta &0 \\
       (c_n-c_t)\cos\delta \sin \delta& \sin^2\delta c_t+ (1+\cos^2\delta)c_n & 0\\ 0 & 0 & c_n  
    \end{array} \right)
    , \;\; 
    \Bo{B}_{tot,p}^{(1)} =\left(
    \begin{array}{ccc}
      0 &0 & -b c_n \sin \delta \\
       0 & 0 & b c_n (1-\cos\delta)\\  
       b c_n \sin \delta  & -b c_n (1-\cos\delta) & 0
    \end{array} \right)
    \\[20pt]
    \Bo{C}_{tot,p}^{(1)} = -
    \left(
    \begin{array}{ccc}
      (1 \!+\!\cos^2\delta)d_t \!+\! \sin^2\delta (d_n\!+\!b^2 c_n) &(d_n\!-\!d_t\!+\!b^2 c_n)\cos\delta \sin \delta & 0 \\
      (d_n\!-\!d_t\!+\!b^2 c_n)\cos\delta \sin \delta  & (1\!+\!\cos^2\delta)(d_n\!+\!b^2 c_n) \!+\! d_t\sin^2\delta  & 0\\  
      0  & 0 & 2(d_n\!+\!b^2 c_n) 
    \end{array} \right)
    \end{array} \]
    \caption{\label{tab.R2} Blocks of the resistance matrices of the ``rigidized'' two-link swimmer}
\end{table*}
One can see that whenever the two links are not aligned $\delta\ne0$, the matrix $\Bo{B}$ of rotation-translation coupling does not vanish. This coupling enables propulsion of the nano-swimmer which is induced by pure external torque. 

\newpage 


\end{document}